\documentclass[pdftex,superscriptaddress,aps,onecolumn]{revtex4}
\usepackage{multirow,enumerate,epsfig,graphics,amssymb,amsmath,subeqnarray,mathrsfs,color,xcolor}
\usepackage{color,epsfig,graphics,amssymb,amsmath,subeqnarray,graphicx,amsthm,subfigure,mathrsfs}
\usepackage[colorlinks=true, citecolor=red, linkcolor=blue ]{hyperref}

\newcommand{\dd}{\mathrm{d}}

\newcommand{\nb}{\mathbf{n}}

\newcommand{\ub}{\mathbf{u}}
\newcommand{\eb}{\mathbf{e}}

\newcommand{\rb}{\mathbf{r}}

\let\grad\nabla
\let\grad\nabla

\newcommand{\pard}[2]{\frac{\partial #1}{\partial #2}}

\newcommand{\bs}{\boldsymbol}

\newcommand{\change}[1]{{#1}}
\newcommand{\pe}{\mbox{Pe}}
\newcommand{\re}{\mbox{Re}}

\begin{document}
\title{Collisions and rebounds of chemically-active droplets}
\author{Kevin Lippera}
\author{Matvey Morozov}
\author{Michael Benzaquen}
\author{S\'ebastien Michelin}
\email{sebastien.michelin@ladhyx.polytechnique.fr}
\affiliation{LadHyX -- D\'epartement de M\'ecanique, Ecole Polytechnique -- CNRS, Institut Polytechnique de Paris, 91128 Palaiseau, France.}
\date{\today}

\begin{abstract}
Active droplets swim as a result of the nonlinear  advective coupling of the distribution of chemical species they consume or release with the Marangoni flows created by their non-uniform surface distribution.  Most existing models focus on the self-propulsion of a single droplet in an unbounded fluid, which arises when diffusion is slow enough (i.e. beyond a critical P\'eclet number, $\pe_c$). Despite its experimental relevance, the coupled dynamics of multiple droplets and/or  collision with a wall remains mostly unexplored.  Using a novel approach based on a moving fitted bispherical grid, the fully-coupled nonlinear dynamics of the chemical solute and flow fields are solved here to characterise in detail the  axisymmetric collision of an active droplet with a rigid wall (or with a second droplet). The dynamics is strikingly different depending on the convective-to-diffusive transport ratio, $\pe$: near the self-propulsion threshold (moderate $\pe$), the rebound dynamics are set by chemical interactions and are well captured by asymptotic analysis; in contrast, for larger $\pe$, a complex and nonlinear combination of hydrodynamic and chemical effects set the detailed dynamics, including a  closer approach to the wall and  a velocity plateau shortly after the rebound of the droplet. The rebound characteristics, i.e. minimum distance and duration, are finally fully characterised in terms of $\pe$.

\end{abstract}
\maketitle

\section{Introduction}
Understanding the fundamental principles and detailed mechanisms of self-propulsion at the microscopic scales has fascinated many researchers across disciplines over the last fifty years~\citep[][]{berg1993,brennen1977,lauga2009}. Biologically, micro-organisms such as bacteria, algae and other moving cells or cell colonies represent a major part of the biomass of our planet, and the individual and collective swimming dynamics  of many of them are critical for the development and reproduction of larger organisms~\citep[][]{suarez2006}, or the dynamics of entire ecosystems~\citep[][]{kirchman2008,guasto2012}. More recently, self-propulsion at such small scales where viscosity completely dominates the effect of inertia, has received much interest for its potential bio-medical or industrial applications, in order to perform particular targeted tasks~\citep[][]{medina2015cellular,park2017multifunctional, singh2017microemulsion}. The collective motion of biological or synthetic microswimmers is also now regarded by physicists as a prototypical example for understanding self-organisation and collective dynamics of individual agents as colonies, swarms and other clusters, and recent years have seen the rapid development of intense research in the field of active matter~\citep{marchetti2013}. Beyond the understanding of the locomotion of a single organism or system, a critical question lies in the interaction of swimmers with their neighbours and environment (e.g. boundaries) to unravel the origin of their collective behaviour~\citep{bechinger2016}.

Developing self-propelled synthetic systems in the lab typically follows a bio-mimetic approach~\citep{dreyfus2005,ghosh2009}, e.g. by mimicking the rotation of chiral filaments or beating of flexible appendages as exploited by many biological species to create motion in the Stokesian realm~\citep{brennen1977}. Yet, in order to overcome the shortcomings or difficulties inherent to this biomimetic approach (e.g. miniaturisation or reliance on a directional macroscopic forcing), a second route has emerged in the last decade relying on the local conversion by active colloids of \change{physico-chemical} energy present in their immediate environment, in order to generate a flow forcing. Such ``catalytic" swimmers can be broadly classified into two different categories, namely autophoretic particles and active droplets.

The former are rigid particles and exploit the general principles of phoresis in order to gain motility, i.e. the generation of a slip flow at a rigid surface under the effect of a gradient in solute concentration, electric potential or temperature \citep{anderson1989colloid,Moran17}; the latter, which are the focus of the present work, are liquid droplets and ``swim" as a result of the Marangoni effect in the presence of a gradient of temperature or surfactant distribution~\citep{Maass16}. In both cases, this mobility is coupled to an activity, namely the ability for the artificial swimmer to generate on its own the \change{physico-chemical} gradients used for propulsion, for example by catalysing a solute-releasing reaction on its surface or through dissolution~\citep{duan2015,Herminghaus14}. The combination of these two fundamental properties provide these systems with the ability to move in a viscous fluid, without relying on externally-controlled and directed fields. As such their dynamics and trajectories are determined solely by their own properties and that of their immediate environment, hence somewhat mimicking the behaviour of living organisms.

In recent years, the spontaneous propulsion of immersed droplets has been reported in numerous experiments~\citep{Maass16,ryazantsev2017thermo}. While the origin of the mobility property of such droplets has clearly and uniformly been identified to Marangoni stresses resulting from local \change{physico-chemical} gradients~\citep{anderson1989colloid}, the origin of the droplet's activity is more elusive and depends strongly on the experimental system considered. \citet{toyota2006listeria} reported the self-propulsion of oil droplets immersed in a micellar medium as a result of the release of large vesicles at their back. Activity may also be tied to a chemical reaction within the droplet altering the nature of the surfactant coverage~\citep{thutupalli2011swarming}. More recently, somewhat simpler systems have been proposed and characterised, that do not rely on a chemical reaction, but rather on the solubilisation of the droplet into the surrounding fluid~\citep{izri2014self,Kruger16,Moerman17}.

In contrast with phoretic ``Janus'' particles whose directionality is directly encoded in their design~\citep{ebbens2016,Moran17}, an active droplet must break a directional symmetry in order to generate \change{physico-chemical} gradients \emph{along} its surface and self-propel in a particular direction. The finite influence of advection by the Marangoni flows appears to be playing a key role here as it introduces a nonlinear feedback of the generated flow field on the distribution of the \change{physico-chemical} property that it results from. This strong coupling, identified early on   at the origin of the instability of sedimenting droplets~\citep{rednikov1995role}, is responsible for the emergence of an instability of the isotropic state of the active droplet and a transcritical bifurcation~\citep{izri2014self,Morozov19a}. This bifurcation and instability are generic to both particles and droplets alike provided diffusion is slow enough compared with the convective transport of solute~\citep[][]{Michelin13b}. These studies identified a self-propulsion threshold in terms of Marangoni or P\'eclet numbers quantifying this advection-to-diffusion ratio, beyond which a self-propelled steady state develops~\citep[][]{schmitt2013swimming,yoshinaga2012drift}. Further transitions to more complex trajectories and chaotic behaviour was also observed experimentally~\citep{Suga18} and analysed theoretically~\citep{Morozov19b,Morozov19c}.

Most models available so far for both phoretic particles and active droplets consider a single micro-swimmer in an unbounded fluid medium (i.e. far from any confining boundary). Yet, most experiments involve many swimmers. Furthermore, the density of the particles or droplets does not match that of the surrounding fluid, and as a consequence, many if not most of them swim close to a bottom rigid wall or a free surface~\citep{theurkauff2012,palacci2013, kruger2016dimensionality}. This interaction and collective dynamics of multiple swimmers is the focus of an increasing attention from the modelling point of view to understand the formation of clusters of particles~\citep{saha2014clusters,pohl2014dynamic}, in particular as a result of the multiple interaction routes available~\citep{Liebchen19,Varma19} or of the effect of the walls on their interactions~\citep{Kanso19,thutupalli2018flow}.  The dynamics of a rigid phoretic particle close to a rigid wall has become a canonical problem to analyse such interactions and the  resulting complex dynamics~\citep{crowdy2013wall,ibrahim2016walls, uspal2015self,yariv2016wall}. It was also shown that interaction and self-assembly of active but individually non-motile particles may also lead to self-propulsion at the collective level~\citep{soto2015self,varma2018clustering}.

In most experimental systems, the chemical dynamics leading to the self-propulsion of phoretic particles is dominantly diffusive, and most of the models discussed above exploit the resulting linearity of the underlying Laplace and Stokes' problems. However, advection and the non-linear coupling it introduces between the chemical and hydrodynamic fields, play  a critical role in the emergence of self-propulsion for active droplets and thus can not be simply neglected. Yet, accounting for this full non-linear coupling in a model or a numerical simulation is no easy task. Several studies have attempted to model the interactions of active droplets, at least within a simplifying limit. \citet{Moerman17} focused on a purely diffusive limit with no hydrodynamic interactions. \citet{yabunaka2016collision} considered the influence of both chemical and hydrodynamic interactions during the collision of two self-propelled droplets, but the approach, which relies on the linear superposition of the hydrodynamic and chemical signature of each droplet, is intrinsically limited to the case of far-field interactions (i.e. when the relative distance of the droplets is large compared to their radii) and to the vicinity of the self-propulsion threshold.   Numerically, \citet{fadda2017lattice} proposed a simulation of the collision problem using a Lattice-Boltzmann framework focusing on the velocity field generated by the two droplets; yet, the solute chemical dynamics and its coupling to the flows it produces, as well as the impact of the proximity of the two droplets, remain elusive at this point.

In contrast with existing modeling efforts on the interaction of two self-propelled droplets (or the interaction of a droplet with a confining wall), the present work aims at the full description of the nonlinearly coupled hydrodynamic and chemical dynamics involved during a head-on (normal) collision. Our approach takes advantage of the axisymmetric setting of the problem but does not require any restrictive assumption regarding either the relative distance of the droplet and the wall (or between the droplets),  the origin of the Marangoni flow which is entirely driven by solute concentration gradients at the droplet's surface or the magnitude of the convective transport with respect to diffusion (quantified in the following by a finite P\'eclet number $\pe$). Our goal is twofold: (i) \change{to} provide an in-depth physical insight into the chemical and fluid dynamics involved during the interaction, in particular to understand how the relative magnitude of advection and diffusion may modify or condition the droplets' collision and rebound; (ii)~\change{to} establish a benchmark study for the collision dynamics, to which reduced models used to analyse the collective behaviour of many droplets could be \change{compared} and validated. 

\begin{figure}
\begin{center}
\includegraphics[width=.7\textwidth]{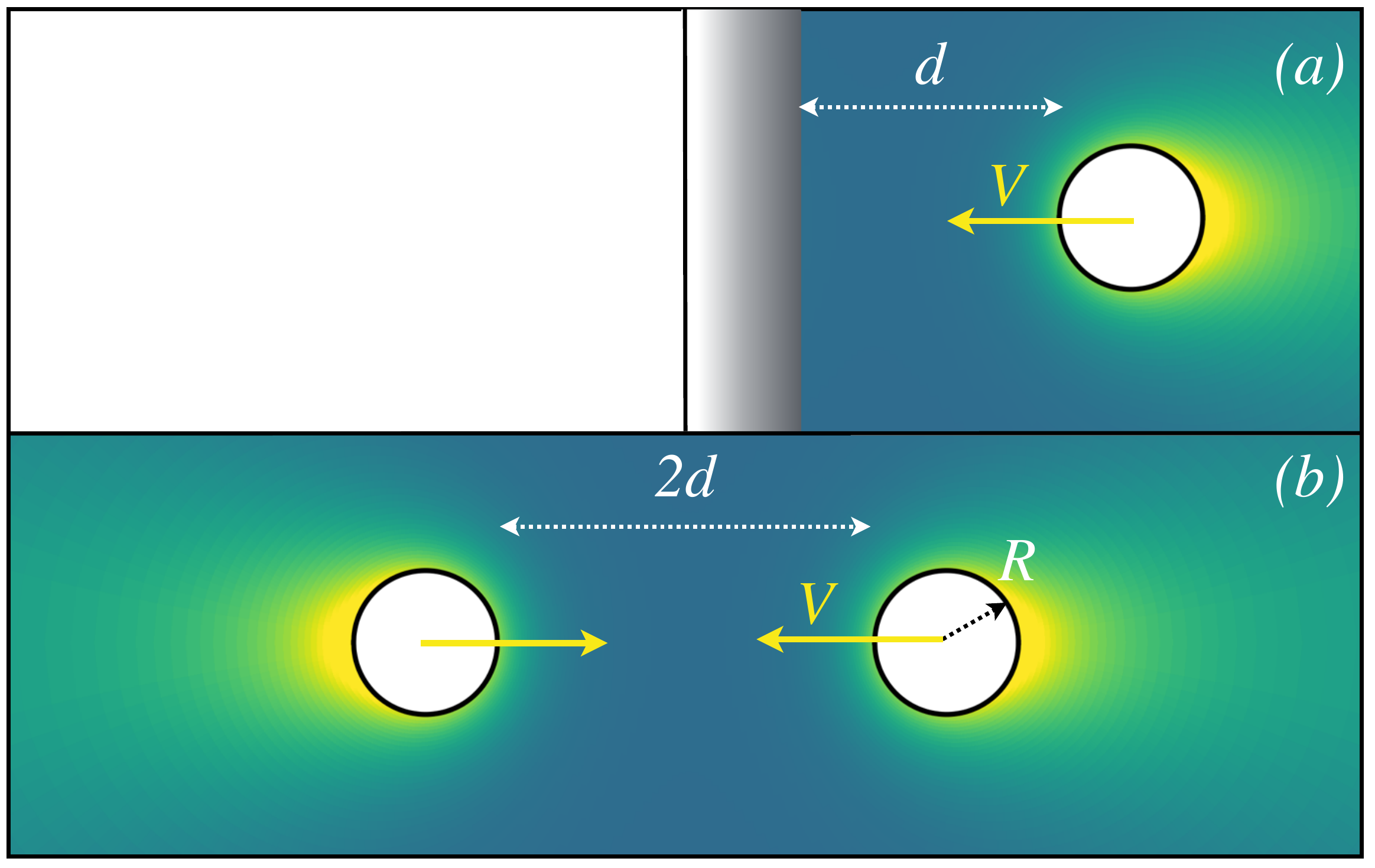}
\caption{$(a)$: Collision between an active droplet and a rigid wall.  $(b)$: Collision between two active droplets. }
\label{schema_physical_problem}
\end{center}
\end{figure}

To this end, we develop a novel framework to analyse the unsteady dynamics of non-linearly coupled hydrodynamic and \change{physico-chemical} systems using a semi-analytical treatment of both problems using bispherical harmonic decompositions on a moving conformal grid, which could be used for the treatment of more generic problems (e.g. bubble dynamics). For simplicity and clarity, because of the strong physical and mathematical similarity between the droplet-droplet and droplet-wall collisions, we focus specifically in the following on the latter problem (i.e. the canonical droplet-wall interaction, Figure~\ref{schema_physical_problem}a) before extending our simulation framework and results to the droplet-droplet collision (Figure~\ref{schema_physical_problem}b, \S~\ref{twodropcollision}).

The paper is organised as follows. Section~\ref{models} introduces the coupled hydrodynamic and chemical problems involved in the interaction, together with the governing equations. In \S~\ref{numericalframework}, our novel treatment of the problem using a spectral decomposition of the fields onto a moving bispherical grid is presented. This approach is used in \S~\ref{DropletInteraction} to analyse in detail the interaction and rebound of a droplet onto a rigid wall, above the self-propulsion threshold, and the different behaviours observed depending on the advection-to-diffusion ratio. The results for a droplet-droplet collision are also presented and discussed. To provide further insight into the behaviour of the system in the vicinity of the self-propulsion threshold (i.e. for small velocity magnitude), a rigorous asymptotic treatment of the interaction is proposed in \S~\ref{asymp}. Our numerical and asymptotic results are then used to propose a quantitatively-accurate effective model of the rebound in \S~\ref{Effective-interaction}, and our findings are finally summarised and discussed in \S~\ref{conclusion}. 

\section{Modelling the interaction between an active droplet and a rigid wall}
\label{models}

\subsection{Physical problem}
We consider a \change{spherical} active droplet of radius $R$ composed of a Newtonian fluid of viscosity $\eta_i$ and density $\rho_i$ swimming normally to a rigid wall ($z=0$) with velocity $\bs V=V(t) \bs e_z$ in a second Newtonian fluid of viscosity $\eta_o$ and density $\rho_o$. The droplet is chemically-active, i.e. it exchanges a chemical solute with the external fluid, whose molecular diffusivity as well as local and background concentrations are noted respectively $D$, $C(\rb,t)$ and $C_\infty$. 

In experiments, two main routes have been identified for this chemical activity \citep{Maass16,Herminghaus14}: (i) a \emph{micellar} pathway, in which empty micelles in the continuous (outer) phase capture some of the inner fluid when approaching the droplet's boundary to form swollen micelles \citep{izri2014self} and (ii) a \emph{molecular} pathway, in which suspended surfactant molecules in the continuous phase incorporate some of the droplet's fluid to form swollen micelles \citep{Moerman17}.
In the following, inspired by the former mechanism, this chemical activity is generically represented as the net release of a chemical solute (e.g. swollen micelles) from the droplet's surface into the outer fluid \change{with a fixed and constant chemical flux per unit time and area,} $\mathcal A>0$ \citep[note that a similar model is obtained from the molecular pathway by considering a fixed capture rate of a surfactant species,][]{Morozov19a}.

\change{Additionally}, the presence of the \change{solute} modifies the local \change{tension of the droplet's interface} and for small enough concentration differences, $\gamma=\gamma_0+\gamma_1 C$ where $\gamma_0$ and $\gamma_1$ are two positive constants \citep{izri2014self,Morozov19a}: surface gradients  of the solute concentration therefore lead to Marangoni stresses at the droplet's interface.

 The solute transport dynamics, which involves both molecular diffusion and advection by the self-generated Marangoni flows, plays therefore a key role in understanding the self-propulsion of a single droplet but also its interaction with a fixed solid wall (Figure~\ref{schema_physical_problem}a) or a second droplet (Figure~\ref{schema_physical_problem}b). In the following, and unless stated otherwise, we focus primarily on the former problem (droplet-wall interaction), and briefly analyse the latter in Section~\ref{twodropcollision}. We denote by $d$ the distance at a given time between the droplet's surface and the wall (Figure~\ref{schema_physical_problem}).

 \subsection{Non-dimensional equations and boundary conditions}
As self-propulsion arises from the Marangoni flows resulting from solute concentration gradients, a natural velocity scale for the problem is given by the drift velocity of a chemically-passive droplet in an externally-imposed gradient of concentration $\mathcal{A}/D$, i.e. $V^*=\mathcal{A}R\gamma_1 /[D(2\eta_o+3\eta_i)]$ \citep{anderson1989colloid}.  In the following, the problem and all quantities of interest are made dimensionless by choosing $R$, $V^*$, $R/V^*$ and $\mathcal{A}R/D$ as characteristic length, velocity, time and concentration scales, respectively.

Noting $\ub^{o,i}(\rb,t)$ the Eulerian velocity field in the outer and inner fluids measured in the fixed laboratory frame, the evolution of the dimensionless relative concentration \change{$c=(C-C_\infty)D/(\mathcal A R)$} in the outer phase follows an advection-diffusion equation:
\begin{align}
\label{diff-adv}
\mbox{Pe}\left(\frac{\partial c}{\partial t}+ \bs u^o \cdot \bs \nabla c\right)=\nabla^2 c\qquad \textrm{with       }\quad\pe=\frac{V^*R}{D}=\frac{\mathcal{A}R^2\gamma_1}{(2\eta_o+3\eta_i)D^2}\cdot
\end{align}
A critical parameter in our study is the P\'eclet number, $\pe$, which physically quantifies the relative importance  of Marangoni advection and molecular diffusion in the solute transport dynamics\change{, and can range between $O(1)$ and $O(10^3)$ depending on the particular physical system considered in experiments~\citep{izri2014self,Moerman17,Suga18}.} 

\change{The droplet's chemical activity is modelled here as a solute emission, which takes the form of a fixed chemical flux}  and we further assume that the rigid wall is unable to exchange any solute with the fluid. Boundary conditions for the relative concentration $c$ are therefore obtained as 
\begin{align}
\label{fixedflux}
\mathbf{n}\cdot\left.\bs \nabla c\right|_{S}=-1,\qquad\left.\mathbf{n}\cdot\bs \nabla c\right|_{W}=0,\qquad
\left.c\right|_{r\rightarrow \infty}=0,
\end{align}
where $S$ (resp. $W$) denotes the surface of the droplet (resp. the wall) and $\mathbf{n}$ the unit normal vector pointing into the outer fluid domain.

Experimentally, the typical radius of active droplets are of the order of $10$--$100\,\mu$m, and their characteristic self-propulsion velocities are $\sim\,10\,\mu$m.s$^{-1}$. At such scales, inertia is negligible (i.e. characteristic Reynolds number are $\re=\rho_oV^*R/\eta_o\sim10^{-3}$) and the flow velocity and pressure satisfy a steady Stokes equation in both phases:
\begin{align}
\label{StokesEq}
\bs \nabla^2 \bs u^{o,i}=\bs \nabla p^{o,i},\qquad\qquad \bs\nabla\cdot\bs u^{o,i}=0.
\end{align}
In the labframe, $\bs u^o$ vanishes far away from the droplet. At the droplet's interface, the velocity field is continuous and Marangoni stresses result in a jump in tangential hydrodynamic stresses. Additionally, a no-slip condition is imposed at the wall surface. Noting $\bs\sigma^{o,i}$, the Newtonian stress tensor in each fluid and $\tilde{\eta}=\eta_i/\eta_o$ the viscosity ratio, the boundary conditions for the hydrodynamic problem are therefore obtained as
\begin{align}
\left.\bs u^o\right|_{S}=\left.\bs u^i\right|_{S},&\qquad \left.(\mathbf{I}-\nb\nb)\cdot (\bs\sigma^o-\tilde{\eta}\bs\sigma^i)\right|_{S}\cdot \mathbf{n} =-(\mathbf{I}-\nb\nb)\cdot\left(2+3\tilde{\eta}\right)\left.\bs \nabla c\right|_S,\label{bcdrop}\\
&\left. \bs u^o\right|_{W}=\bs 0,\qquad\left.\bs u^o\right|_{r\rightarrow \infty}=\bs 0, \label{bcwall}
\end{align}
and recalling that $\bs V$ is the droplet's surface velocity, the impermeability condition further imposes
\begin{equation}
\left.(\bs u^o \cdot \mathbf{n})\right|_{S}=\bs V\cdot\mathbf{n},\label{imperm}
\end{equation}
It should be noted here that we assume that the droplet remains spherical, since the mean surface tension $\gamma_0$ is typically much larger than characteristic hydrodynamic stresses, $\eta_oV^*/R$ \change{, i.e. the typical capillary number, $\mbox{Ca}=\eta_o V^*/R\approx 10^{-4}$, is much lower than unity \citep[][]{izri2014self}}.  Neglecting inertia, the droplet must remain force-free which provides an additional implicit relation to determine $\bs V$:
 \begin{align}
 \label{forcefree}
\bs F=\int_{S}\bs \sigma^o \cdot \mathbf{n} \,\mathrm{d}S=\bs 0.
\end{align}
Note that due to the axisymmetric nature of the problem, the droplet is necessarily also torque-free.

\subsection{Droplet velocity and polarity}
Once $c$ is known, the hydrodynamic problem is linear and is essentially similar to any surface-driven swimming problem. The total hydrodynamic force $\bs F$ on the droplet can therefore be decomposed into two distinct parts: (i) a Marangoni force $\bs F_m$ experienced by a \emph{fixed} droplet (i.e. $\bs V=0$) in response to the concentration distribution at its surface, and (ii) a hydrodynamic drag $\bs F_d$  resulting from its translation and obtained from the classical  Hadamard-Rybczynski problem of a translating droplet with uniform surface tension~\citep{Leal07}. Due to the linearity of the latter \change{problem with respect to the droplet velocity}, $\bs F_d=-\bs R\cdot\bs V$ \change{where $\bs R$, the resistance matrix,} is a function only of the geometry of the problem (i.e. the droplet radius and its distance to the wall). \change{Using  $\bs F=\bs F_d+\bs F_m$, Eq.~\eqref{forcefree} becomes},
\begin{align}
\change{\bs V=\bs R^{-1}\cdot\bs F_{m}}.
\end{align}

While the hydrodynamic problem \emph{for fixed $c$} is linear, the overall dynamics of the solute concentration and droplet motion are not. Indeed, advection by the self-induced Marangoni flows introduces a nonlinearity which is key to understand the self-propulsion of an isotropic droplet or chemically-active particle~\citep{Michelin13b}: when diffusion is slow enough, advection maintains a front-back polarity of the concentration gradient that drives the necessary Marangoni flows. In other words, beyond a critical P\'eclet number \change{($\pe_c=4$)} the isotropic state of a single active droplet becomes unstable and self-propulsion can develop~\change{\citep{izri2014self,Morozov19a}}.

It is therefore clear that the front-back asymmetry of the concentration field plays a critical role in the self-propulsion dynamics and in the following, this asymmetry is characterised by the polarity $\bs \Pi$ \change{of the surface concentration},
\begin{equation}
\bs \Pi =-\frac{1}{2\pi}\int_{S}c\mathbf{n} \mathrm{d}S.\label{eq:PiDef}
\end{equation}
\change{In the case of a single droplet in an unbounded fluid, Lorentz' Reciprocal Theorem for Stokes flow can be used to obtain the droplet velocity directly in terms of the surface tension (or solute concentration) gradient~\citep{pak2014,masoud2019}:
\begin{equation}
\bs V=-\frac{1}{4\pi}\int_S(\mathbf{I}-\nb\nb)\cdot\grad c\,\dd S,
\end{equation}
which demonstrates, after integration by part, that the droplet's velocity and chemical polarity match exactly for a single droplet:
}
\begin{eqnarray}
\bs V=\bs \Pi.\label{VPi}
\end{eqnarray}
\change{It should be noted that this result stems purely from hydrodynamics, and therefore does not depend on the solute dynamics (or the P\'eclet number $\pe$). In the axisymmetric configuration considered here, both velocity and polarity are along $\eb_z$ and we thus focus in the following on their respective axial projections $V=\bs V\cdot\mathbf{e}_z$ and $\Pi=\bs\Pi\cdot\eb_z$}

\section{Solving for the coupled hydrodynamic and chemical fields}\label{numericalframework}
The physical effect of the wall on the approaching droplet is \emph{a priori} two-fold: hydrodynamically, the confinement of the fluid between the droplet and the bounding wall modifies the viscous stresses (and droplet's resistance matrix $\bs R$); furthermore, the chemically-inert wall reduces the effective solute diffusion away from the droplet resulting in an accumulation of solute in front of the approaching droplet. Because the active droplet is anti-chemotactic (it swims down the solute concentration gradient), the latter effect is expected to repel the droplet.

To account in details for these two effects and their coupling, a novel analytical and numerical framework is proposed and detailed here to solve exactly for the non-linearly coupled dynamics of the flow field and solute advection-diffusion using a moving bispherical coordinate system matching the moving droplet's boundary. 

\subsection{Bi-spherical coordinates system}
\label{adv-diff-bisph}
\begin{figure}
\centering
\includegraphics[width=.5\textwidth]{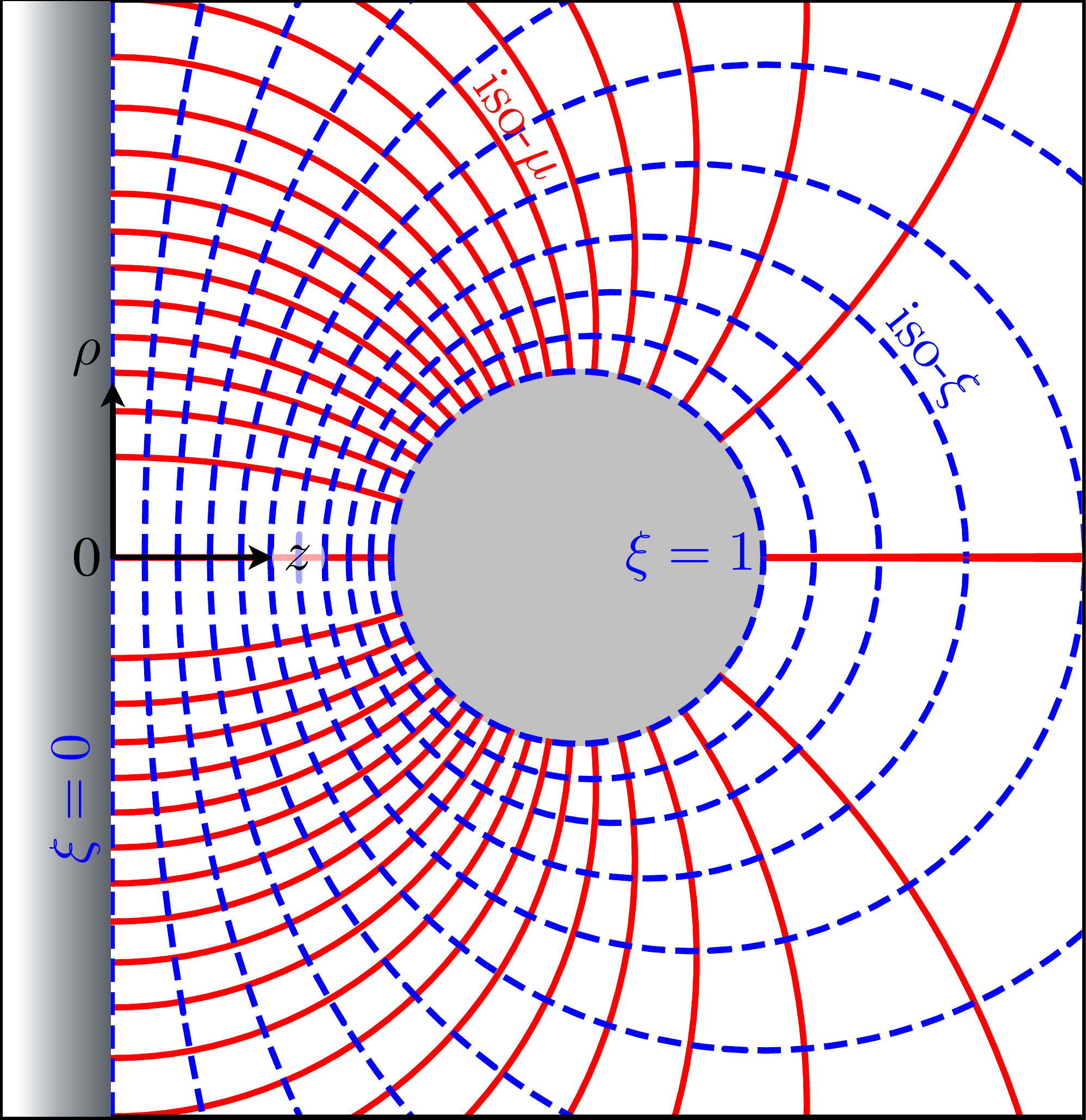}
\caption{Bi-spherical coordinate system. Contours of fixed $\mu$ (solid red) and fixed $\xi$ (dashed blue) are shown at a given time $t$. The surface of the wall and droplet are given by $\xi=0$ and $\xi=1$, respectively.}
\label{bisphcoords}
\end{figure}
Cartesian coordinates are well-adapted to describe fluid motion or solute transport above a flat wall, yet spherical coordinates are typically more convenient to describe the flow and solute dynamics near \change{the} droplet surface. A body-fitted mesh is thus defined to describe simply both boundaries using bispherical coordinates, an approach that is convenient to apply boundary conditions (Figure~\ref{bisphcoords}). In contrast with many studies using such coordinates~\citep{stimson1926motion,popescu2011,michelin2015autophoretic,reigh2015}, the droplet is not fixed with respect to the wall so that the bispherical system needs to be modified at each time to match the evolving boundaries.

At a given time $t$, a point located at $(z,\rho,\phi)$ in the fixed cylindrical coordinate system (with $\eb_z$ the axis of symmetry of the problem and the origin located on the wall) has bispherical coordinates $(\xi,\mu,\phi)$ defined by:
\begin{equation} \label{eq:bispher_coord}
\rho=\frac{a(t)\sqrt{1-\mu^2}}{\Gamma},\qquad z=\frac{a(t)\sinh(\lambda(t)\xi)}{\Gamma}\qquad \textrm{with   }\Gamma(\xi,\mu,t)=\cosh(\lambda(t)\xi)-\mu,
\end{equation} 
where $a(t)$ and $\lambda(t)$ are functions of time to account for the time-dependent stretching of the grid. Surfaces of constant $\xi$ represent a set of non-intersecting spheres (Figure~\ref{bisphcoords}). At any time $t$, the wall and droplet's surfaces correspond to $\xi=0$ and $\xi=1$, respectively. The functions  $a(t)$ and $\lambda(t)$ are determined uniquely from the droplet's radius and distance to the wall:
\begin{equation}
\lambda(t)=\cosh^{-1}\left(\frac{d(t)}{R}+1\right),\qquad a(t)=R\sqrt{d(t)(d(t)+2R)}.
\end{equation}
In addition, the unit vectors of the bispherical basis are defined as $(\eb_\xi,\eb_\mu,\eb_\phi)$ with
\begin{eqnarray}
\label{basis1}
\bs e_{\xi}&=& \frac{1-\mu\cosh(\lambda\xi)}{\Gamma}\bs e_z  -\frac{\sqrt{1-\mu^2}\sinh(\lambda\xi)}{\Gamma} \bs e_{\rho},\\
\label{basis2}
\bs e_{\mu}&=& \frac{\sqrt{1-\mu^2}\sinh(\lambda \xi)}{\Gamma} \bs e_z +  \frac{1-\mu\cosh(\lambda\xi)}{\Gamma}\bs e_{\rho}.
\end{eqnarray}
and the corresponding metric coefficients are
\begin{equation}\label{metric}
h_{\xi}=\frac{a\lambda}{\Gamma},\qquad h_\mu=\frac{a}{\Gamma\sqrt{1-\mu^2}},\qquad
h_{\phi}=\frac{a\sqrt{1-\mu^2}}{\Gamma}\cdot
\end{equation}

\subsection{Grid adaptation for unsteady problems}
Because of the motion of the droplet (and resulting grid adaption), a point of fixed $(\xi,\mu,\phi)$ is not fixed in the labframe, i.e. it has time-dependent $(\rho,z)$-coordinates. This has consequences when solving time-dependent equations such as Eq.~\eqref{diff-adv}. Indeed, considering the local change in time of the concentration field $c$ at a \emph{fixed} point ($\rho,z$) now introduces a material derivative when considering $c$ as a function of $(\xi,\mu,t)$, and we must thus replace: 
\begin{eqnarray}
\left.\frac{\partial c }{\partial t}\right|_{\rho,z}=\left.\frac{\partial c }{\partial t}\right|_{\xi,\mu}-\bs\chi\cdot\bs\nabla c
\end{eqnarray}
where $\bs\chi$ is the velocity of a point with fixed $(\xi,\mu)$ in the physical space, and is obtained from Eq.~\eqref{eq:bispher_coord}. The advection-diffusion equation for $c(\xi,\mu,t)$ is therefore obtained as
\begin{equation}
\left.\frac{\partial c}{\partial t}\right|_{\xi,\mu} +\left(\bs u-\bs\chi\right) \cdot \bs \nabla c=\frac{1}{\mbox{Pe}}\nabla^2 c,
\label{newdiffadvequation}
\end{equation}
where, noting time derivatives of single-variable functions with a dot symbol, 
\begin{equation}
\bs\chi \cdot \bs \nabla c =\left(\frac{\dot{\lambda}\xi}{\lambda}-\frac{\dot{a}\mu\sinh(\lambda\xi)}{\lambda a }\right)\frac{\partial c}{\partial \xi}+\frac{\dot{a}}{a}(1-\mu^2)\cosh(\lambda\xi)\frac{\partial c}{\partial \mu}.\quad\quad\quad
\label{chidef}
\end{equation}

\subsection{Hydrodynamical problem}\label{sec:hydro}
Solving the transport equation \eqref{newdiffadvequation} requires knowing the velocity field $\bs u$. As noted before, the hydrodynamic (Stokes) problem is instantaneous and linear and the classical method to obtain Stokes flow solutions in bispherical geometries can be used. The (inner and outer) flow fields are obtained in terms of streamfunctions $\psi^{i,o}$ \citep{michelin2015autophoretic}:
\begin{equation}
u^{i,o}_{\xi}= \frac{\Gamma^2}{a^2} \frac{\partial \psi^{i,o}}{\partial \mu},\qquad 
u^{i,o}_{\mu}= -\frac{\Gamma^2}{a^2\lambda\sqrt{1-\mu^2}} \frac{\partial \psi^{i,o}}{\partial \xi},
\label{psi-u}
\end{equation}
which can be written for an axisymmetric problem as~\citep{stimson1926motion}:
\begin{equation}
\label{psiLegendre}
\psi^{i,o}(\xi,\mu,t)=\Gamma^{-3/2}\sum_{n=1}^{\infty}(1-\mu^2)L_n'(\mu)U_n^{i,o}(\xi,t),
\end{equation}
where $L_n'$ is the first derivative of $L_n$ the Legendre polynomial of degree $n$, and the functions $U_n^{i,o}$ are given by
\begin{align}
U^o_{n}(\xi,t)=&\ \alpha_n\cosh\left[\left(n+\frac{3}{2}\right)\lambda\xi\right]+\beta_n\sinh\left[\left(n+\frac{3}{2}\right)\lambda\xi\right]+\gamma_n\cosh\left[\left(n-\frac{1}{2}\right)\lambda\xi\right]\nonumber\\
&+\delta_n\sinh\left[\left(n-\frac{1}{2}\right)\lambda\xi\right],\\
U^i_{n}(\xi,t)=&\ \tilde{\alpha}_ne^{-(n+3/2)\lambda|\xi|}+\tilde{\beta}_ne^{-(n-1/2)\lambda|\xi|},
\end{align}
where $\alpha_n$, $\beta_n$, $\gamma_n$, $\delta_n$, $\tilde{\alpha}_n$ and $\tilde{\beta}_n$ are determined independently at each instant $t$ from the kinematic and dynamic boundary conditions on the droplet and the wall surfaces. The continuity and impermeability conditions at the droplet's boundary, Eqs.~\eqref{bcdrop} and \eqref{imperm}, become using both \change{\eqref{basis1} and \eqref{psi-u}}: 
\begin{equation}
\label{impermeabilitybisph0}
\left.\frac{\partial \psi^i}{\partial \mu}\right|_{\xi=1}=\left.\frac{\partial \psi^o}{\partial \mu}\right|_{\xi=1}=\frac{a^2(1-\mu\cosh\lambda)}{(\cosh\lambda-\mu)^3}V.
\end{equation}
Integrating with respect to $\mu$ along the droplet's boundary (and imposing $\psi^i=\psi^o=0$ on the axis of symmetry which is a streamline of the problem) one obtains:
\begin{eqnarray}
\left.\psi^i\right|_{\xi=1}=\left.\psi^o\right|_{\xi=1}=\frac{(1-\mu^2)a^2}{2(\cosh\lambda-\mu)^2}V.
\label{impermeabilitybisph}
\end{eqnarray}
The continuity of the velocity field at the droplet's boundary further imposes
\begin{equation}\label{continuity2}
\left.\frac{\partial \psi^o}{\partial \xi}\right|_{\xi=1}=\left.\frac{\partial \psi^i}{\partial \xi}\right|_{\xi=1},
\end{equation}
and the Marangoni condition in Eq.~\eqref{bcdrop} at the surface of the droplet becomes:
\begin{eqnarray}
\left.(\sigma^o_{\xi\mu}-\tilde\eta\sigma^i_{\xi\mu})\right|_{\xi=1}=-\left.\frac{(2+3\tilde\eta)(\cosh\lambda-\mu)\sqrt{1-\mu^2}}{a} \frac{\partial c}{\partial \mu}\right|_{\xi=1}.
\label{marangonibisph}
\end{eqnarray}
Finally, the no-slip boundary condition at the wall, Eq.~\eqref{bcwall}, becomes:
\begin{equation}
\label{noslip}
\left.\psi^o\right|_{\xi=0}=0,\qquad\left.\frac{\partial \psi^o}{\partial \xi}\right|_{\xi=0}=0.
\end{equation}
Equations~\eqref{impermeabilitybisph}--\eqref{noslip} projected in the polar direction along the $n$-th Legendre polynomial (Appendix~\ref{ProjHyd}) provide $n$ sets of 6 linear equations, that determine  $\alpha_n$, $\beta_n$, $\gamma_n$, $\delta_n$,  $\tilde{\alpha}_n$ and $\tilde{\beta}_n$ (and thus the streamfunction) uniquely in terms of the surface concentration distribution.

\subsection{Transport problem}
Solving the transport equation, Eq.~\eqref{newdiffadvequation}, also exploits a spectral decomposition of $c$ along the Legendre polynomials in the polar direction. Inspired by the separated form of the solution for Laplace's equation in bispherical coordinates, the relative concentration field $c$ (which vanishes at infinity here) is thus decomposed as
\begin{eqnarray}
\label{cLegendre}
c(\xi,\mu,t)&=&\Gamma^{1/2}\sum_{n=0}^{\infty}c_n(\xi,t)L_n(\mu),
\end{eqnarray}
where the $c_n(\xi,t)$ functions are yet to be determined. \change{Using Eqs.~\eqref{metric} and \eqref{psi-u}, the advection-diffusion equation, Eq.~\eqref{newdiffadvequation}, becomes:}
\begin{align}
\left.\frac{\partial c}{\partial t}\right|_{\xi,\mu}&+\left[ \frac{\dot{a}\mu\sinh(\lambda\xi)}{\lambda a}-\frac{\dot{\lambda}\xi}{\lambda}+\frac{\Gamma^3}{\lambda a^3}\frac{\partial \psi^o}{\partial \mu}\right]\frac{\partial c}{\partial \xi}-\left[\frac{\dot{a}}{a}(1-\mu^2)\cosh(\lambda\xi)+\frac{\Gamma^3}{\lambda a^3}\frac{\partial \psi^o}{\partial \xi} \right]\frac{\partial c}{\partial \mu}\nonumber\\
&=\frac{1}{\mbox{Pe}}\frac{\Gamma^3}{\lambda a^2}\left[\frac{1}{\lambda}\frac{\partial}{\partial \xi}\left(\frac{1}{\Gamma}\frac{\partial c}{\partial \xi}\right) + \frac{\partial}{\partial \mu}\left(\frac{(1-\mu^2)}{\Gamma}\frac{\partial c}{\partial \mu}\right) \right],
\end{align} 
and substituting \change{$\psi^{o}$} and $c$ from Eqs.~\eqref{psiLegendre} and \eqref{cLegendre} yields:
\begin{align}
\label{modal_adv_diff}
&\sum_{n=0}^{\infty}\Bigg\{\frac{L_n}{\Gamma^{1/2}}\pard{c_n}{t}+\frac{\dot a}{a}\left[\frac{(1+\mu\cosh(\lambda\xi))L_n-2\cosh(\lambda\xi)(1-\mu^2)L_n'}{2\Gamma^{1/2}}\right]c_n\nonumber\\
&+\left(\frac{\dot{a}\mu\sinh(\lambda\xi)-\dot\lambda\xi a}{\lambda a}\right)\frac{L_n}{\Gamma^{1/2}}\pard{c_n}{\xi}+\frac{1}{\lambda a^3}\sum_{k=1}^\infty\left[\left(\frac{3}{2}(1-\mu^2)L_k'L_n-k(k+1)\Gamma L_kL_n\right)U_k\pard{c_n}{\xi}\right.\nonumber\\
&+\left.\frac{\lambda\sinh(\lambda\xi)}{2}\left[3(1-\mu^2)L_k'L_n'- k(k+1)L_nL_k\right]U_kc_n+(1-\mu^2)L_k'\left(\frac{L_n}{2}-\Gamma L_n'\right)\pard{U_k}{\xi}c_n\right]\Bigg\}\nonumber\\
&\qquad=\frac{\Gamma^{3/2}}{a^2\mbox{Pe}}\sum_{n=0}^{\infty}\bigg(\frac{1}{\lambda^2}\frac{\partial^2 c_n}{\partial \xi^2}-\left(n+\frac{1}{2}\right)^2c_n\bigg)L_n.
\end{align}
Projecting the advection-diffusion equation, Eq.~\eqref{modal_adv_diff}, onto $L_p(\mu)$ provides a set of coupled \change{partial differential equations} for $\bs C(\xi,t)=[c_0(\xi,t),c_1(\xi,t),...,c_N(\xi,t)]$  (see Appendix \ref{OperatorProjection}) which can be formally written as:
\begin{align}
\bs H\cdot \frac{\partial \bs C}{\partial t}&+\left(\bs B^{1}\cdot \bs U+\bs B^{2}\cdot \frac{\partial \bs U}{\partial \xi} + \bs G^1\right) \cdot \bs C\nonumber\\
&+\left(\bs B^{3}\cdot \bs U + \bs G^2\right) \cdot \frac{\partial \bs C}{\partial \xi}=\frac{1}{\text{Pe}}\left(\bs A^1\cdot \bs C+ \bs A^2\cdot \frac{\partial^2 \bs C}{\partial \xi^2} \right),
\label{mainPDE}
\end{align}
where the second-order tensors $\bs H$, $\bs A^i$, $\bs G^i$ ($i=1,2$) and the third-order tensor $\bs B^{j}$ ($j=1,2,3$) have coefficients that are obtained in terms of integrals of appropriate combinations of Legendre polynomials and depend on $\xi$ (see Appendix ~\ref{OperatorProjection}). Physically, terms in $\bs H$, $\bs G^i$, $\bs B^j$ and $\bs A^i$ are related to the local time-derivative (for fixed $\xi$ and $\mu$), the grid adaptation, convection by the Marangoni flow and diffusion, respectively. In Eq.~\eqref{mainPDE}, $\bs U(\xi,t)=[U_1^o(\xi,t),U_2^o(\xi,t),...,U_N^o(\xi,t)]$ is a linear and instantaneous function of $\bs C$ (see \S~\ref{sec:hydro}).

Finally, this set of evolution equations for $c_n(\xi,t)$, Eq.~\eqref{mainPDE}, which are second-order in space, must be complemented by appropriate boundary conditions. After substitution of Eq.~\eqref{cLegendre}, and projection onto the $n$-th Legendre polynomial the flux conditions, Eqs.~\eqref{fixedflux}, at the droplet's ($\xi=1$) and wall's ($\xi=0$) surfaces  become:
\begin{align}
\label{BCC1}
 \left.\left(\frac{\lambda\sinh \lambda}{2} \,c_n+\cosh\lambda \frac{\partial c_n}{\partial \xi}-\frac{n+1}{2n+3}\frac{\partial c_{n+1}}{\partial \xi}-\frac{n}{2n-1}\frac{\partial c_{n-1}}{\partial \xi}\right)\right|_{\xi=1}&=\sqrt{2}a\lambda e^{-(n+1/2)|\lambda|},\\
 \label{BCC2}
\left.\left(\frac{\partial c_n}{\partial \xi}-\frac{n+1}{2n+3}\frac{\partial c_{n+1}}{\partial \xi}-\frac{n}{2n-1}\frac{\partial c_{n-1}}{\partial \xi}\right)\right|_{\xi=0}&=0.
\end{align}

At each time-step, the functions $(U^o_n)_n$ are obtained from $(c_n)_n$ following the results of \S~\ref{sec:hydro}. Note that only the outer flow needs to be known explicitly. The set of non-linear \change{partial differential equations}, Eq.~\eqref{mainPDE}, is then solved using finite differences and a uniform grid of $N_\xi$ points with $0\leq \xi\leq 1$. Advective terms, i.e. those involving $U_n^o(\xi)$, are treated explicitly while a Crank-Nicholson scheme is used to account for the diffusive terms.

It should be noted here that the present framework can easily be adapted to consider the collision of two droplets rather than a droplet with a wall: for two symmetric droplets, the $\xi$-grid is simply extended to $-1\leq \xi\leq 1$ with the boundary conditions at the wall ($\xi=0$) being replaced by appropriate conditions on the second droplet ($\xi=-1$). Alternatively, for a purely symmetric situation (see Section~\ref{twodropcollision}), we only need to solve for the right-half plane and the right-most droplet by imposing the symmetry of the concentration and velocity field on $\xi=0$.

\subsection{Validation of the numerical model}
\begin{figure}
\centering
\includegraphics[width=.8\textwidth]{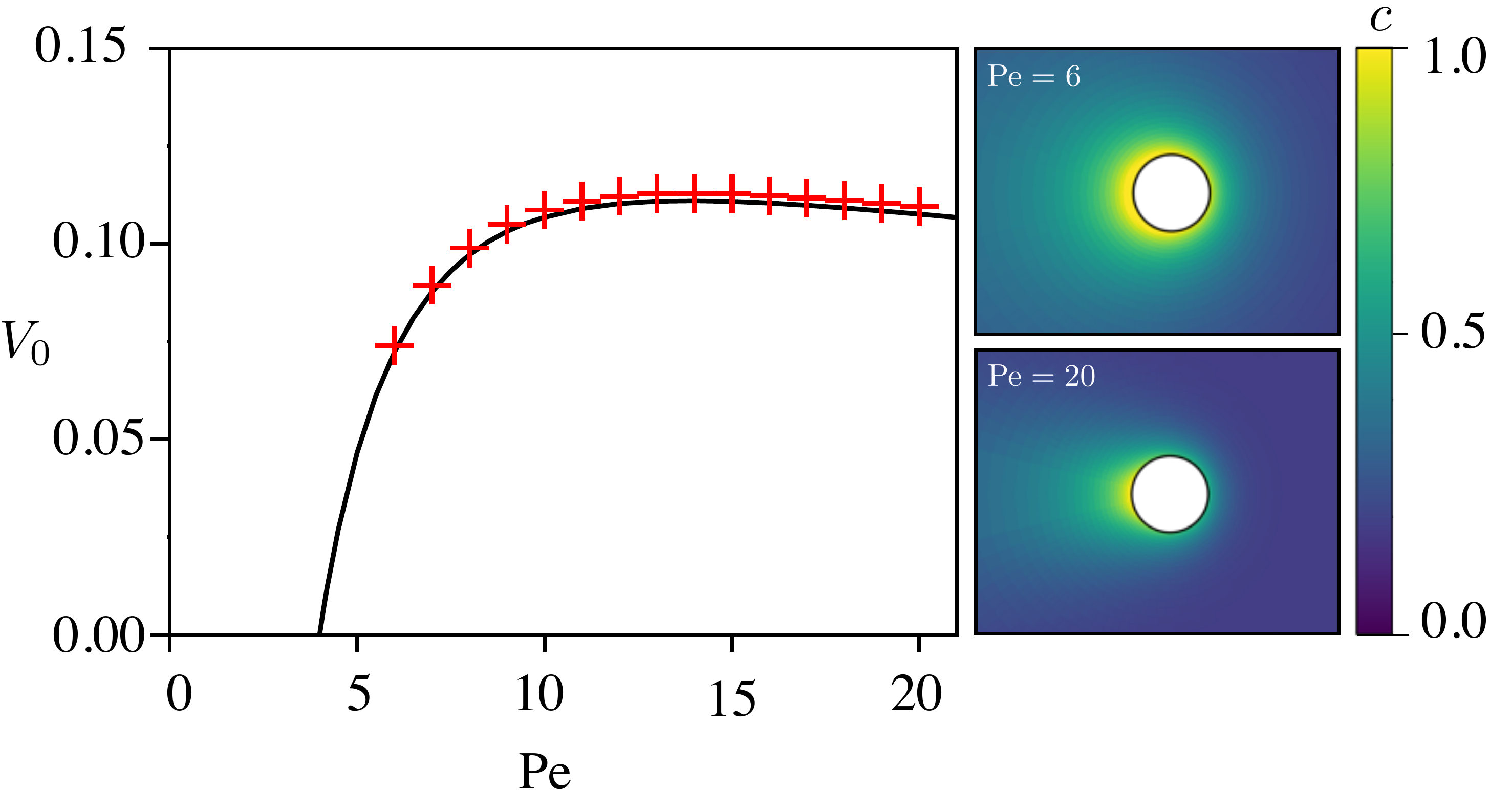}
\caption{Mean self-propulsion velocity $V_0$ of two active droplets with $\tilde\eta=1/36$  initially located at a distance $d=48$ from each other  (red crosses). The results of Ref.~\citep{izri2014self} for a single active droplet with the same viscosity ratio are also reported (black solid line)\change{, and the concentration distribution around the self-propelling droplets is shown for $\pe=6$ and $\pe=20$.}  }
\label{izricompare}
\end{figure}

The validity and accuracy of the present approach are tested by comparison with the case of a single self-propelled droplet considered in Ref.~\citep{izri2014self}. To this end,  we consider the case of two droplets initially separated by a distance $d=48$, i.e. far enough that one expects their interaction to be only weak and to recover the single-droplet results. The concentration field is initialised using the purely-diffusive ($\pe=0$) solution for which an analytical solution is available for $c$ \citep{michelin2015autophoretic}. At $t>0$, the previous simulation framework is used for a fixed non-zero value of $\pe$; both droplets are initially forced to move at a fixed positive velocity $V=0.1$ until $t=2$ and are let to evolve force-free for $t>2$; after a transient regime, their velocity relaxes toward a fixed and common value identified as their self-propulsion velocity $V_0$ when isolated. The mean long-time velocity of the droplets is measured at $t=3000$; it is reported on Figure~\ref{izricompare} and compared to the results of \citet{izri2014self} for a single droplet. The results are in excellent agreement and validate the present framework: the maximum relative errors obtained are around $2\%$ for $6\leq\pe\leq 20$, when using  $N=60$ polar modes and $N_\xi=100$ regularly-spaced grid points in $\xi$, a resolution precise enough to guarantee the accurate description of the physical processes, yet light enough to analyse the time-dependent rebound dynamics of the droplet as well as the influence of $\pe$ on the detailed chemo-hydrodynamic interaction between the droplet and the wall. 

\section{Droplet interaction with a rigid wall}
\label{DropletInteraction}
The main focus of this section is the collision of a self-propelled droplet with a rigid and passive wall. For simplicity, the inner and outer fluid viscosities are assumed identical from now on ($\tilde\eta=1$). The motion of the droplet as well as the front-back asymmetry of the concentration field, i.e. the motion's primary driving mechanism, are monitored by the axial droplet's velocity $\bs V$ and the polarity of the surface concentration $\bs \Pi$ defined in Eq.~\eqref{eq:PiDef}, respectively. After analysing the collision dynamics and how it is influenced by the advection-to-diffusion ratio $\pe$, the case of the symmetric collision of two identical droplets is also briefly considered for comparison.

\subsection{Collision dynamics for moderate Pe}
\begin{figure}
\centering
\includegraphics[width=.9\textwidth]{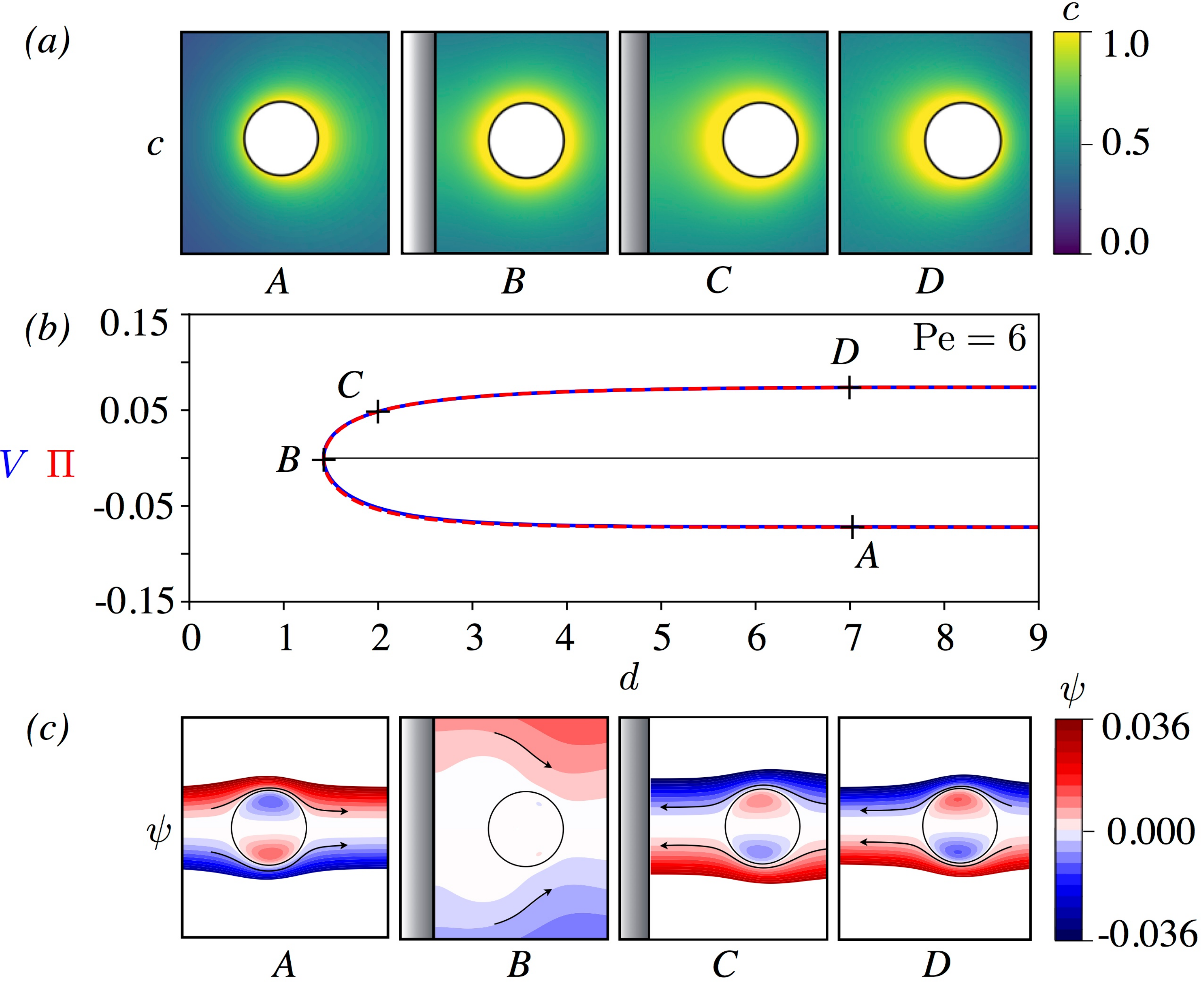}
\caption{Collision of an active droplet with a rigid wall at $\pe=6$. \change{The evolution of the droplet velocity $V$ (blue solid) and polarity $\Pi$ (red dashed) during the collision are reported in terms of the distance $d$ between the droplet's surface and the wall (b). Snapshots of the concentration (a) and streamfunction (c) are also shown for four representative positions indicated as (A-D) on panel (b). }}
\label{collision_overviewPe6}
\end{figure}

We first analyse the collision dynamics for moderate advection, i.e. for $\pe$ slightly above the instability threshold ($\pe=6$ is chosen here). Figure~\ref{collision_overviewPe6} displays the evolution of \change{the axial velocity $V$ and polarity $\Pi$ of the concentration field} with the distance, \change{ $d$}, between the droplet and the wall. A corresponding movie (Movie 1) is also provided as Supplementary Material. Initially, the droplet swims toward the wall with a constant self-propulsion velocity $-V_0$, where $V_0(\pe)$ is the magnitude of self-propulsion of an isolated droplet. As it approaches the wall, it decelerates and reverses direction ($V=0$) at a finite distance $d_\textrm{rb}(\pe)$ from the wall ($d_\textrm{rb}=1.4$ for $\pe=6$). In a second phase, the droplet accelerates away from the wall and eventually reaches again its self-propulsion velocity $V_0$. A main observation of Figure~\ref{collision_overviewPe6} is that the velocity and polarity are almost equal throughout the collision and the consequence of this is discussed in more depth below. In the following, we analyse each sequence of the interaction to identify the roles of the different mechanisms.

\subsubsection{Far-field interactions and droplet deceleration}

The droplet is expected to respond to the wall's influence on both the chemical and hydrodynamic fields. Chemically, the droplet acts as a source of chemical. The no-flux boundary condition prevents the diffusion of solute through it which essentially amounts to an elevation of the solute content in the wall's vicinity. When the droplet is far enough from the wall, this amounts to an effective image source of chemical located in the $z<0$ half-plane creating a $1/d^2$ chemical gradient and repulsive Marangoni force on the droplet. When the droplet is close enough, this repulsion eventually dominates the self-propulsion maintained by the chemical polarity at the droplet's surface. 

Hydrodynamically, the wall modifies the drag coefficient on the droplet but also modifies the swimming velocity resulting from a given traction applied at the droplet's surface (here Marangoni stress). Figure~\ref{collision_overviewPe6} shows that the polarity and velocity remain almost identical throughout the collision, as for a single isolated droplet, although both quantities evolve in time due to the modification of the concentration field. The equality of $V$ and $\Pi$ for a single isolated droplet, Eq.~\eqref{VPi}, solely stems from the hydrodynamic problem, which suggests that the hydrodynamic influence of the wall is weak here; in other words, changes in the droplet velocity result mainly from the modification of the concentration distribution at its surface (i.e. chemical interactions) and not from hydrodynamic interactions with the wall which appear subdominant.

\subsubsection{Near-field and re-acceleration toward self-propulsion}
  When the droplet velocity vanishes (\change{instant B in Figure~\ref{collision_overviewPe6}}), the polarity of the concentration field also comes close to zero. A closer look at the distribution of chemical on the surface at that instant in fact reveals that the concentration distribution is almost homogeneous (its variance is reduced by an order of magnitude, when compared to the initial self-propelling state): this results in the droplet's arrest as there is no longer a Marangoni effect acting on the surface of the droplet and the fluid is at rest \change{(see Figure~\ref{collision_overviewPe6})}. 
  
  \begin{figure}[t]
\centering
\includegraphics[width=.75\textwidth]{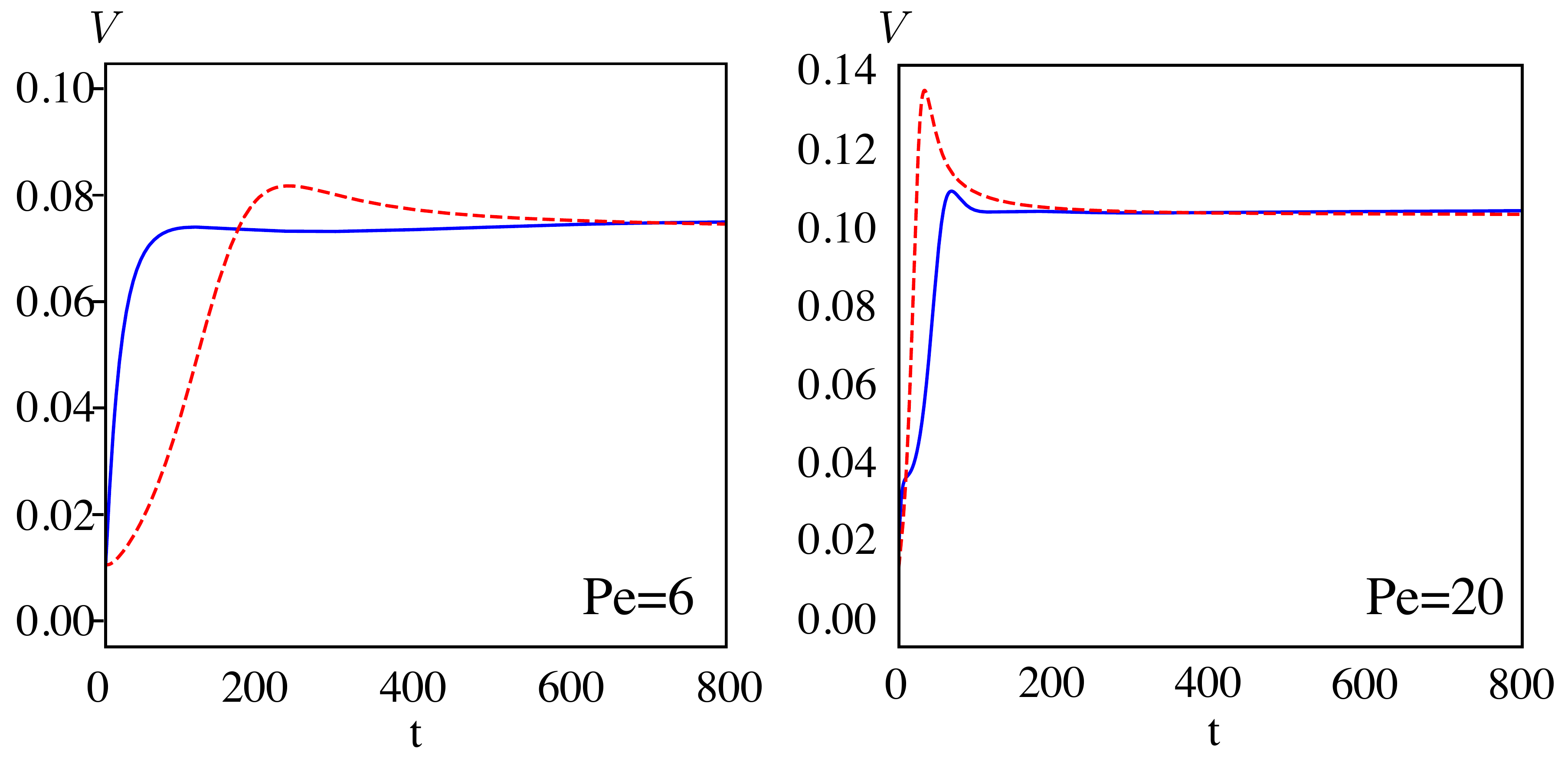}
\caption{Evolution in time of the droplet's velocity for $\pe=6$ (left) and $\pe=20$ (right) during its re-acceleration after its collision with the wall (solid blue). This evolution is compared to the acceleration of an isolated droplet initially forced with a positive velocity $V=0.1$ before being released force-free at $t=0$ (dashed red). }\label{overshootPe6}
\end{figure}

  However, this equilibrium is only ephemeral as the presence of the wall promptly breaks this uniform distribution: the \change{chemical flux at} the droplet's surface being spatially uniform, the confinement on the side of the wall leads to an increased solute concentration there and a repelling Marangoni force. As a result, the droplet drifts away from the wall (Figure~\ref{collision_overviewPe6}, instant C). Because $\pe$ is greater than the instability threshold, this perturbation of the concentration field simultaneously leads to the development of the same instability phenomenon that conferred the droplet its initial velocity, until it reaches $V_0$ as the droplet moves far away from the wall. 
  
  In order to study the influence of the wall in the droplet ``forced'' re-acceleration, Figure~\ref{overshootPe6} (left) compares this second phase of the motion with the situation of a single isolated droplet initially pushed at a finite velocity in the positive $z$ direction before left force-free. The acceleration is initially greater in the second part of the droplet collision with the wall than in the reference isolated droplet case. Indeed, the presence of the wall reinforces the droplet acceleration (by accumulating more solute at its back) than in the case where the solute is able to diffuse freely.

\subsection{Collision at higher Pe}

The results obtained for moderate $\pe$ presented a rather simple picture of the collision dynamics: dominated by the chemical interactions with the wall, it amounts to a slowing down and repulsion of the droplet under the effect of the accumulating chemical solute in front of it due the confining presence of the wall, the wall's hydrodynamic effect being mostly subdominant. The picture becomes however much more complex as the importance of convection of solute vs. diffusion is increased, and the focus of this section is to analyse how wall interactions and collision dynamics are modified as $\pe$ is increased.

\begin{figure}
\centering
\includegraphics[width=.9\textwidth]{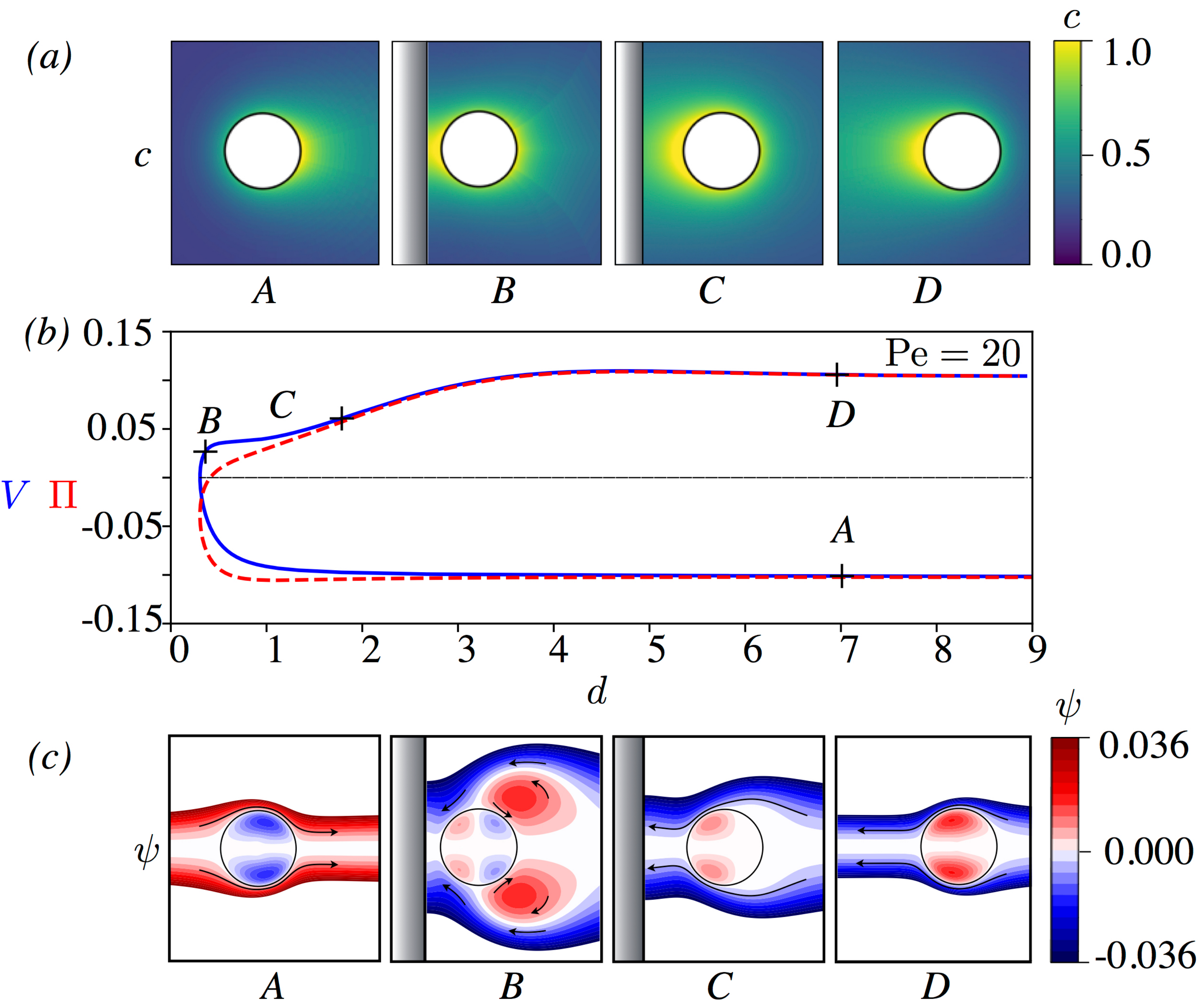}
\caption{Collision of an active droplet with a rigid wall at Pe $=20$. \change{The evolution of the droplet velocity $V$ (blue solid) and polarity $\Pi$ (red dashed) during the collision are reported in terms of the distance $d$ between the droplet's surface and the wall (b). Snapshots of the concentration (a) and streamfunction (c) are also shown for four representative positions indicated as (A-D) on panel (b). } }
\label{collision_overviewPe20}
\end{figure}
Figure~\ref{collision_overviewPe20} presents the evolution of velocity and polarity throughout the collision for $\pe=20$. A corresponding movie (Movie 2) is also provided as Supplementary Material. As for $\pe=6$, the droplet initially propels at $-V_0$ toward the wall (A) and decelerates up to a stopping point(B). This minimum rebound distance, $d_\textrm{rb}$ is however much lower ($d_\textrm{rb}=0.3$ for $\pe=20$). Also, unlike for moderate $\pe$, the re-acceleration of the droplet is not a smooth process. In particular it displays a clear velocity plateau right after the rebound during which the droplet velocity remains almost constant (C). Eventually, and as expected, the droplet reaches once again its self-propulsion velocity as it moves away from the wall whose influence becomes negligible (D). In contrast with the moderate-$\pe$ collision (Figure~\ref{collision_overviewPe6}), we note that the polarity and velocity do not match one another anymore during most of the near-field interactions with the wall, suggesting a stronger hydrodynamic influence of the confinement. 

 \subsubsection{Early interactions and droplet deceleration}
In contrast with the case of moderate advection ($\pe=6$) for which the droplet starts to slow down at a distance $d\approx4.5$ away from the wall, the velocity of the droplet remains relatively \change{unchanged for larger $\pe$} (e.g. down to a distance $d\approx 3$ for $\pe=20$, Figure~\ref{collision_overviewPe20}). This slowing down of the droplet was identified as predominantly associated with the chemical repulsion resulting from the confinement of its own chemical signature. This approach of the droplet closer to the wall is therefore consistent with the faster  (exponential) decay of the concentration field in front of the droplet as a result of the solute advection,  while the decay is only algebraic in its wake~\citep{Acrivos62,Michelin14,Morozov19a}. The asymmetric structure of the concentration field can be observed by comparing \change{instants A on } Figures~\ref{collision_overviewPe6} and~\ref{collision_overviewPe20} \change{(see also Figure~\ref{izricompare})}. As a result, the direct influence of the wall on the concentration field arises belatedly during the interaction for larger $\pe$. 

\begin{figure}
\centering
\includegraphics[width=.75\textwidth]{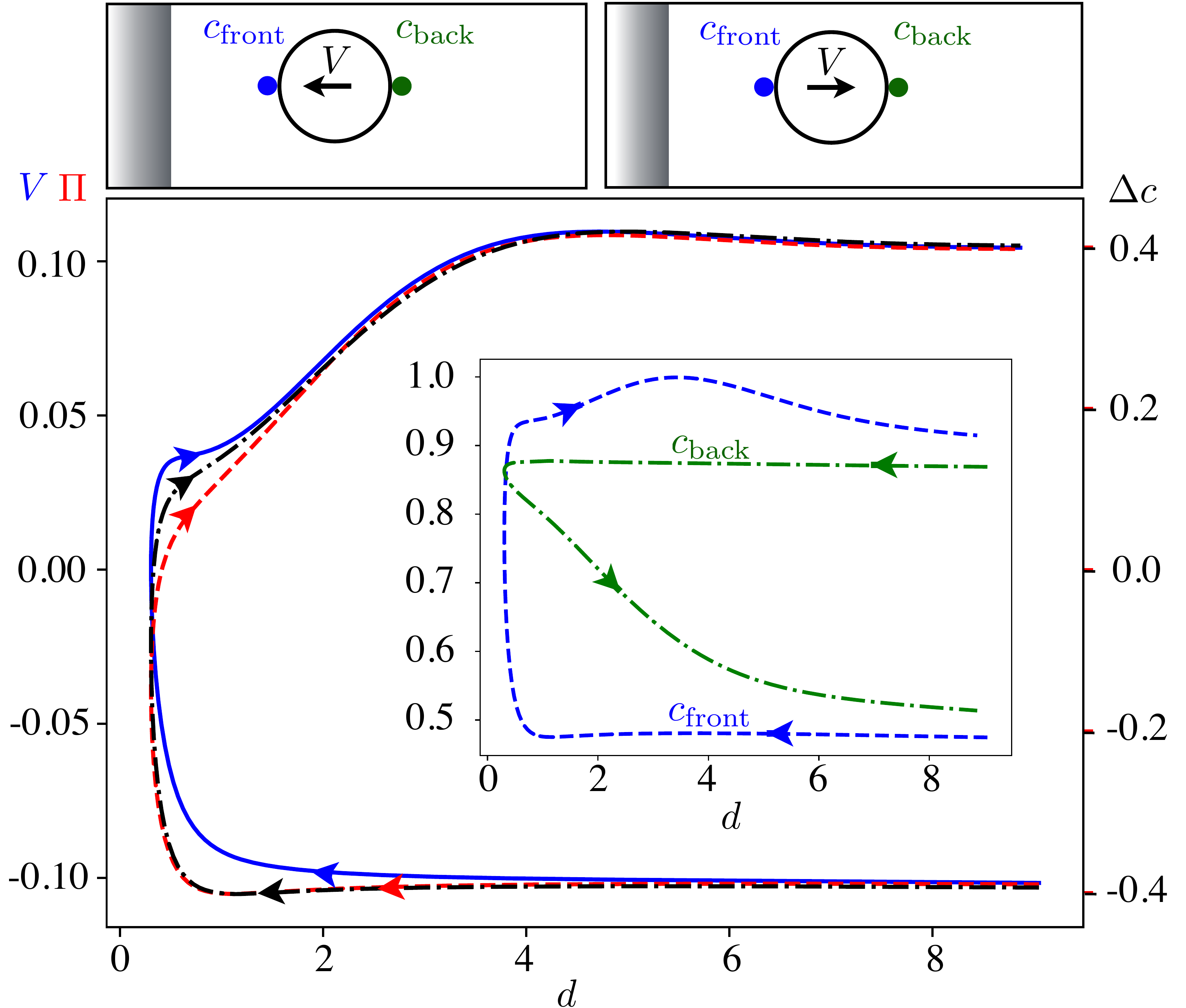}
\caption{Evolution as function of the droplet-wall distance $d$ of the velocity $V$ (solid blue), the polarity $\Pi$ (red dashed) and front-back surface concentration contrast ${\Delta c=c_{\text{front}}-c_{\text{back}}}$ (black dashed-dotted) \change{at $\mbox{Pe}=20$}. \change{The inset shows} the evolution of $c_{\text{back}}$ and $c_{\text{front}}$ with $d$ individually and the definition of these two quantities is reminded at the top of the figure.}
\label{early_chemicalPe20}
\end{figure}

We already noted that $V$ and $\Pi$ do not match one another anymore, contrary to the moderate $\pe$ regime, indicating a direct hydrodynamic influence of the wall. Strikingly, and contrary to the intuition that chemical confinement would reduce the front-back concentration contrast at the droplet's surface, we also note that the polarity of the surface concentration $|\Pi|$ is increased as the droplet approaches the wall, and reaches its maximal value close to $d=1$ before sharply reversing as the droplet stops. To understand this phenomenon in greater depth, the front-back concentration difference at the surface  $\Delta c=c_{\textrm{front}}-c_{\textrm{back}}$ is represented on Figure~\ref{early_chemicalPe20}. Note that $\Delta c$ is a second measure of the asymmetry in surface concentration which evolves in the same manner as $|\Pi|$ during the first part of the motion. During the approach of the droplet, both back and front concentrations are observed to increase (down to a distance $d=3$). In a second phase, $V$ starts to decrease under the effect of hydrodynamic interactions. Yet, the Marangoni flow is not stopped, and in fact contributes to maintain the concentration contrast responsible for a net pumping flow toward the back of the droplet effectively expelling more solute toward the droplet's wake (illustrated by the decrease of $c_{\text{front}}$), which therefore explains the increase of $|\Pi|$.

As in the moderate-$\pe$ case, solute accumulates between the wall and the droplet as they get closer to each other, but the minimum distance $d_{\text{rb}}$ of the droplet surface to the wall is now significantly smaller as a result of the sharper decay of the surface concentration ahead of the droplet during the approaching phase. This induces a sharper increase of the concentration between the wall and droplet, resulting in the fast inversion of $V$ observed in Figure~\ref{collision_overviewPe20}. Figure \ref{early_chemicalPe20} shows how the inversion of $V$ can directly be correlated to the increase of $\Pi$.
\subsubsection{Rebound \& Velocity plateau}
\label{Velocityplateau}

A distinguishing feature of the collision dynamics for higher $\pe$ is the existence, shortly after the rebound of the droplet, of a velocity plateau during which the droplet's velocity remains relatively constant, and significantly smaller than $V_0$, while the droplet moves away from the wall by  about one radius \change{(from B to C on Figure~\ref{collision_overviewPe20})}. To understand its origin, \change{Figure~\ref{collision_overviewPe6}(c) and \ref{collision_overviewPe20}(c) present} the evolution of the flow field (streamlines) in the frame of reference of the droplet for $\pe=6$ and $\pe=20$\change{, respectively}. For $\pe=6$, the flow field is very weak at the instant of rebound (\change{B}), a consequence of the homogeneity of the surface concentration and resulting absence of Marangoni forcing. 

In contrast, at that same instant for $\pe=20$, a strong flow in and around the droplet is observed to persist as a consequence of the surface concentration inhomogeneity (\change{instant B on Figure~\ref{collision_overviewPe20}}). Within our Stokesian approach, this flow field is an instantaneous response to the concentration distribution at the droplet surface. This flow helps sustain the polarity of the arrested droplet while balancing the chemical repulsion introduced by the wall; as a result, a velocity plateau develops until the flow within the droplet reverses \change{(C)}. The structure of the flow within the droplet is quadrupolar (in contrast with the dipolar flow observed during self-propulsion) and is a direct result of the surface concentration distribution, whose slow relaxation for larger $\pe$ introduces a delay before the instability leading to the droplet's self-propulsion away from the wall may develop again.  The evolution of $c_{\text{back}}$ in Figure~\ref{early_chemicalPe20} illustrates the mitigation of the residual amount of solute at the back of the droplet during the second part of the rebound.

\subsubsection{Re-acceleration toward self-propulsion}

Since $d_{\text{rb}}$ is lower at higher $\pe$, one would expect an enhanced repulsion from the wall and therefore an even faster re-acceleration of the droplet (when compared to the development of the self-propulsion instability for an isolated droplet) than was observed for moderate $\pe$ (Figure~\ref{overshootPe6}, left). This is however not the case: strikingly, and in contrast with the moderate-$\pe$ situation, the droplet actually takes more time to recover its propulsion velocity $V_0$ after the rebound than if it was alone (Figure~\ref{overshootPe6}, right). This effect is a direct consequence of the persistence of an excess of solute in the droplet's wake after its approach to the wall, which was already shown to create a pumping flow that holds the droplet back.

\subsection{Rebound distance for varying Pe}
\label{SeveralPe}

\begin{figure}
\centering
\includegraphics[width=.7\textwidth]{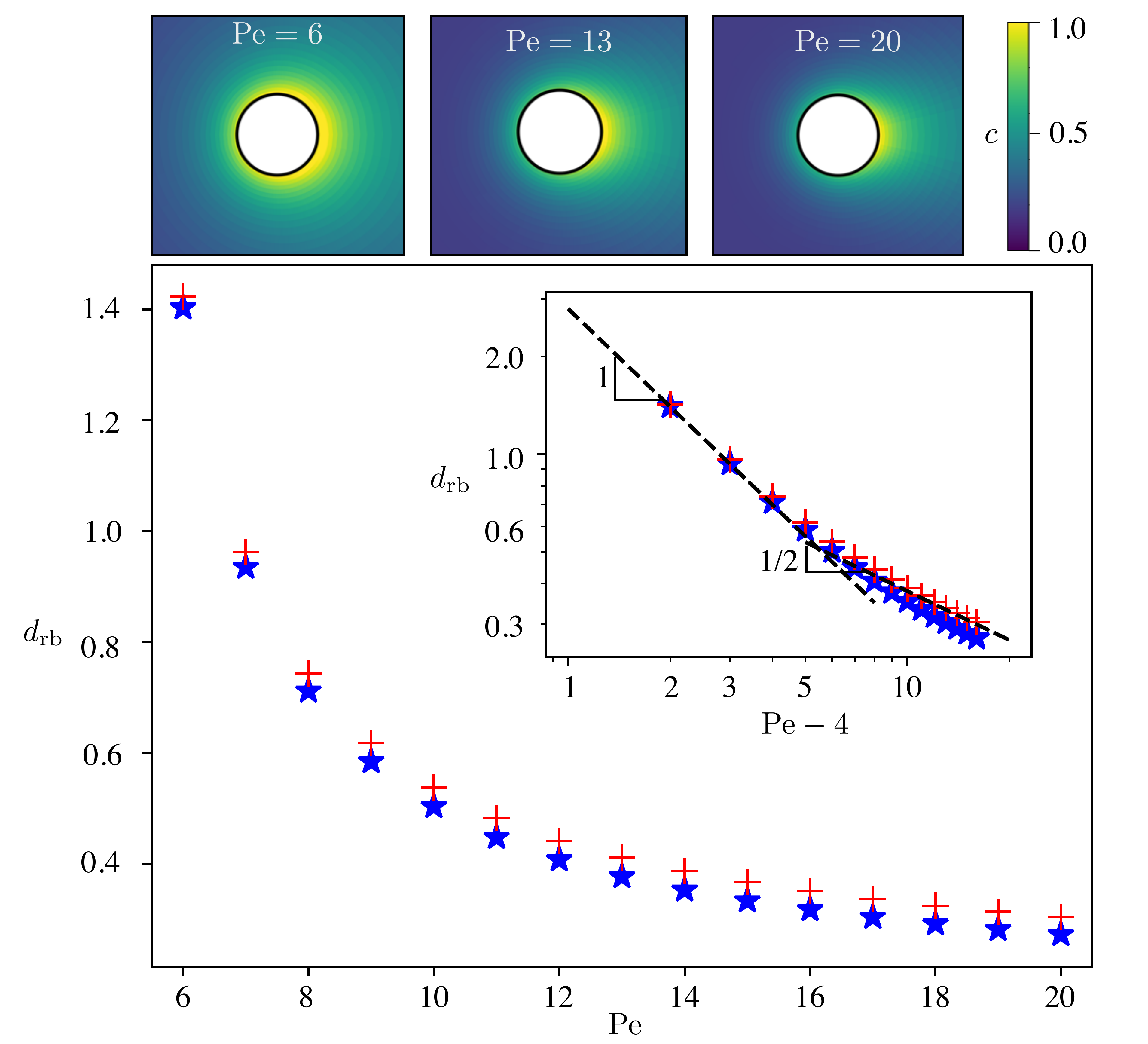}
\caption{\change{Evolution of the rebound distance $d_\textrm{rb}$ (measured when $V=0$) for the droplet-wall collision (red crosses) and the two-droplet collision (blue stars, see \S~\ref{twodropcollision}). The inset }reproduces the same data in log-log scales to identify the asymptotic scaling for near-critical and large $\pe$. The concentration fields around an active droplet approaching the wall (i.e. propelling to the left) are also shown \change{for three representative P\'eclet numbers:} $\pe=6$, $\pe=13$ and $\pe=20$.}
\label{rebound_distance}
\end{figure}

In the previous sections, the droplet-wall interaction  was analysed in details for two different values of $\pe$. One key feature was that, due to the structure of the concentration field ahead of the moving droplet, the rebound distance (i.e. the minimum distance of approach of the droplet to the wall) is reduced when advection plays a more important role in the solute transport. The goal of this section is to provide a more complete characterisation of this phenomenon and we thus now focus on the evolution with $\pe$ of the distance $d_{rb}$ between the wall and the front of the droplet at the time it reverses direction (Figure~\ref{rebound_distance}). A monotonic decrease of $d_\textrm{rb}$ is observed, as expected from the structure of the concentration field ahead of the moving droplet (see top panels of Figure~\ref{rebound_distance}).

We also note  the existence of two distinct regimes in this decrease. For moderate $\pe$, and as $\pe$ approaches the minimum value for self-propulsion \change{($\pe_c=4$)},  $d_\textrm{rb}$ diverges as $(\text{Pe}-\text{Pe}_c)^{-1}$: close to the self-propulsion threshold, the droplet is more sensitive to the wall's influence and is repelled at much greater distances. The asymptotic analysis of the collision near $\pe_c$, presented in \S~\ref{asymp}, confirms this scaling and provides more \change{insight} on the interaction and rebound dynamics. For larger $\pe$ (typically $\pe\gtrsim 10$), a slower decrease is observed as $(\text{Pe}-\text{Pe}_c)^{-1/2}$.

\subsection{Comparison to the two-droplet collision}
\label{twodropcollision}
\begin{figure}
\centering
\includegraphics[width=.7\textwidth]{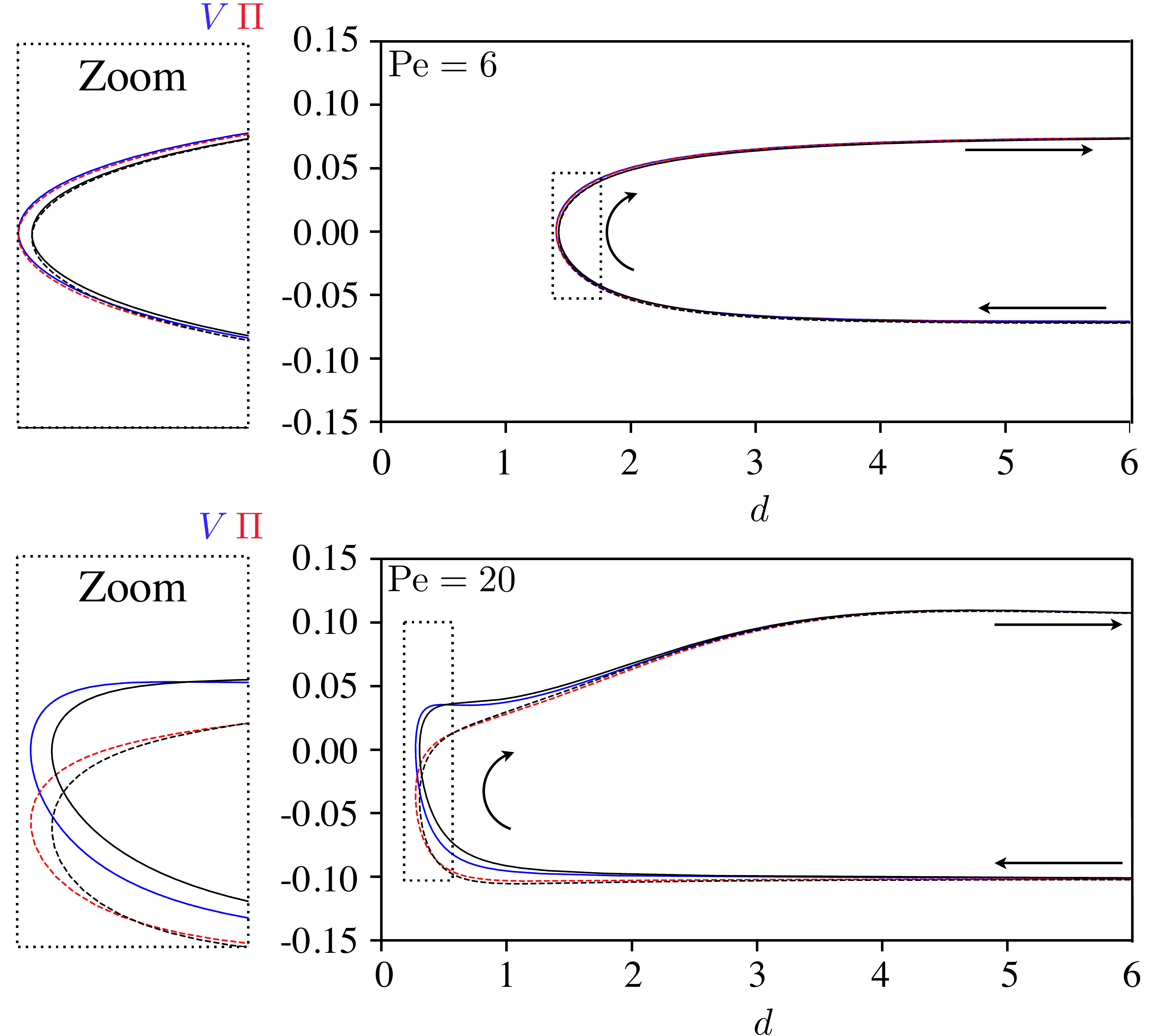}
\caption{Collision between two active droplets at $\pe=6$ (top) and $\pe=20$ (bottom). In each case, the evolution with $d$, the half-distance between the two droplets, of the droplet velocity $V$ (solid blue) and polarity $\Pi$ (dashed red) is shown. For reference the corresponding results for the collision of a single droplet with a no-slip wall are also shown for the velocity (solid black) and polarity (dashed black). }
\label{collision_overview_drop}
\end{figure}

The interaction of a droplet with a chemically-inert wall shares many similarities to the symmetric interaction of two identical droplets. Mathematically, the chemical problem is in fact identical, the no-flux condition at the wall in Eq.~\eqref{fixedflux} being strictly equivalent to a perfect symmetry of the concentration field as for the case of two symmetric droplets. The only difference therefore lies in the hydrodynamic problem and resulting flow field: a no-slip condition is applied for the case of rebound on a wall, Eq.~\eqref{bcwall}, while symmetry conditions on the velocity field would hold on $z=0$ for the case of two droplets (effectively amounting to the presence of a free surface rather than a rigid wall). The modification of the hydrodynamic field may nevertheless have significant consequences, in particular at larger $\pe$ due to the importance of advection by this flow field on the solute dynamics. 

Before closing this section, we therefore briefly analyse the difference between the two situations in more details. Defining $2d$ the minimum distance between the surfaces of the two droplets (Figure~\ref{schema_physical_problem}b), and \change{using} the same numerical approach as for the droplet-wall interaction, we report in Figure~\ref{collision_overview_drop} the evolution with $d$ of the velocity and polarity of the right-hand droplet for moderate and higher $\pe$, and compare those results to that obtained for the wall collision. Both configurations lead to the same dynamics except in the immediate vicinity of the stopping point where some small variations can be identified. These differences are more pronounced for larger $\pe$, which is likely due to the closer proximity to the wall. Indeed, the no-slip condition applied at the wall surface is relaxed for the two droplet collision and therefore the hydrodynamical interactions close to the wall are not identical \citep[see][chap. 12]{kim2013microhydrodynamics}. 

Figure~\ref{rebound_distance} further \change{compares} the evolution of $d_{\text{rb}}$ with $\pe$ for the droplet-wall and droplet-droplet collisions. Once again the evolutions are similar, although the rebound distance is systematically smaller in the case of two droplets (or of a free surface) than for a droplet and a no-slip wall.  This may be due to the different structure of the flow field between the droplet and wall/symmetry plane: a no slip condition on the wall prevents the fluid to flow along the $z=0$ axis and thus can not convect away as much solute as in the case of two droplets.

\section{Asymptotic analysis of the collision dynamics for $\pe\approx\pe_c$}
\label{asymp}

So far we have accrued a general understanding of the droplet-wall and droplet-droplet interactions in the presence of advection. In this Section, we employ asymptotic methods to explain some of our findings in more detail for the symmetric collision of two identical droplets (although most of the reasoning below will be shown to be straightforwardly applicable to the collision of a single droplet with a no-slip wall). Since explicit analytical treatment of advection-diffusion in bispherical coordinates is exceedingly complex, we consider a pair of identical active droplets separated by a large center-to-center distance ${2 d_c \equiv 2 D / \epsilon}$, where $\epsilon \ll 1$. We postulate that the system remains symmetric at all times, i.e., droplets either approach each other with relative velocity $-{2 V}$ or part ways with velocity ${2 V}$. We also assume that the P{\'e}clet number is close to the critical value ${\pe_c = 4}$ corresponding to the spontaneous onset of self-propulsion,
\begin{equation}
  \label{asymp_pe_exp}
  \pe = \pe_c + \epsilon \delta,
\end{equation}
where $\delta = O(1)$ is the supercriticality parameter. In this case, a weakly-nonlinear theory of droplet interaction may be constructed using only axisymmetric spherical coordinates in the vicinity of each droplet. In particular, we will obtain an asymptotic solution to the problem formulated by Eqs.~\eqref{diff-adv}--\eqref{forcefree} by considering each of the droplets separately using axisymmetric spherical coordinates co-moving with the corresponding droplet.

\begin{figure}
\begin{center}
\includegraphics[width=.7\textwidth]{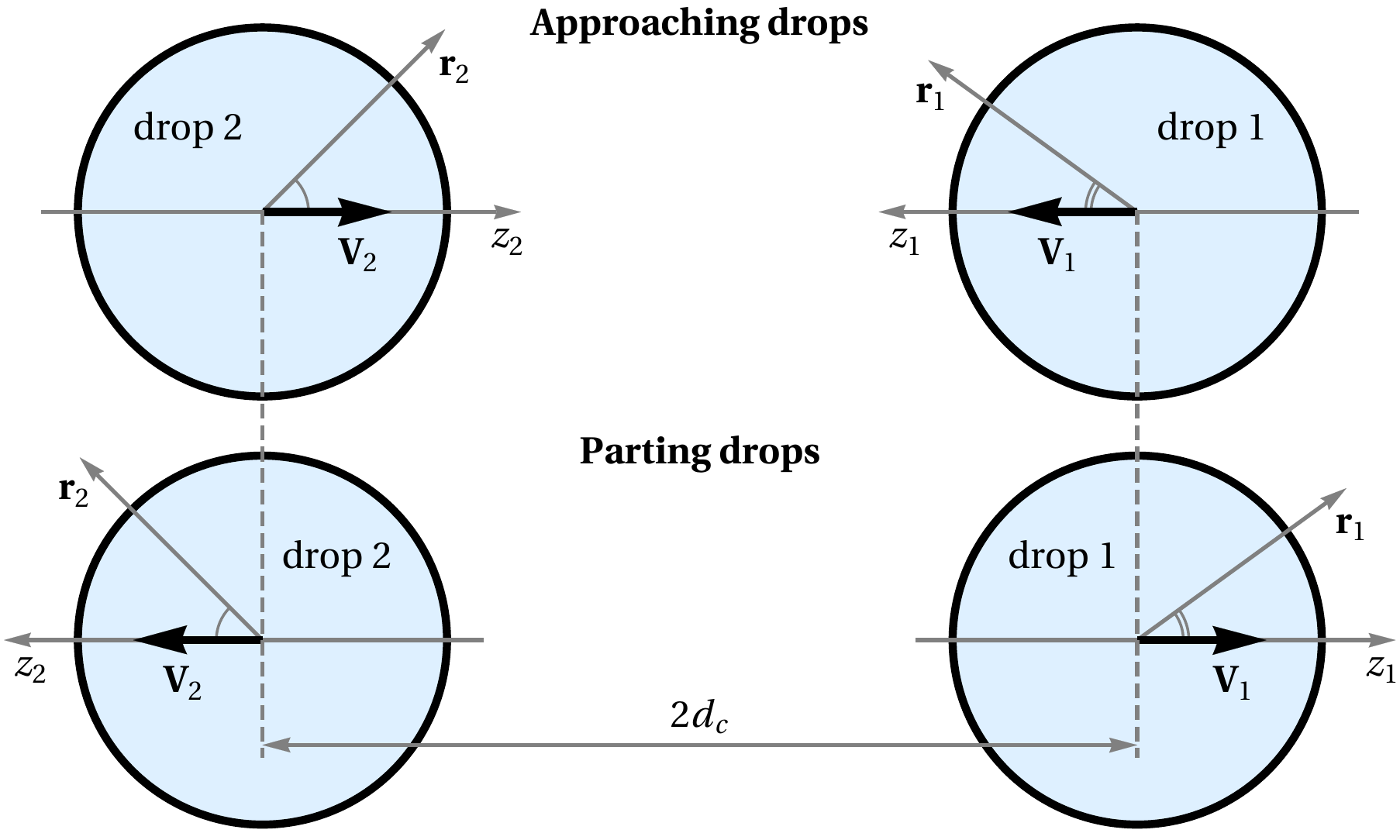}
\caption{Asymptotic analysis and notations for the case of two approaching (top) or departing (bottom) active droplets. The direction of reference of each system of spherical coordinates (i.e. $\theta_i=0$) is given by the swimming direction of the corresponding droplet, and is therefore opposite for the two droplets.}\label{fig_2drops}
\end{center}
\end{figure}

In the vicinity of the self-propulsion threshold (i.e. for $\epsilon\ll 1$), the droplet velocity is expected to be small,
\begin{equation}
  \boldsymbol V_i = \epsilon \boldsymbol V_i^{(1)} 
    + \epsilon^2 \boldsymbol V_i^{(2)} + \ldots
  \quad \text{for  } i = 1,2,
\end{equation}
and, thus, advection is weak~\citep{Rednikov94, Morozov19a}. In the limit of weak advection, the chemical footprint of an individual droplet is known to consist of a near-field part, $N(\textbf{r})$, valid for ${r \ll 1/\epsilon}$, and a far-field contribution, $F(\textbf{r})$, valid for ${r \gg 1}$~\citep{Acrivos62, Rednikov94, Morozov19a}. Accordingly, we seek for a quasi-steady solution of the problem and expand the near- and far-field components of the concentration field of each droplet in powers of $\epsilon$,
\begin{eqnarray}
  \label{asymp_n_exp}
  N_i(\textbf{r}_i) =& N_i^{(0)}(\textbf{r}_i) +& \epsilon N_i^{(1)}(\textbf{r}_i) 
    + \epsilon^2 N_i^{(2)}(\textbf{r}_i) + \ldots
  \quad  \text{for  } i = 1,2, \\
  \label{asymp_f_exp}
  F_i(\boldsymbol \rho_i) =& & \epsilon F_i^{(1)}(\boldsymbol \rho_i)
    + \epsilon^2 F_i^{(2)}(\boldsymbol \rho_i) + \ldots
  \quad  \text{for  } i = 1,2,
\end{eqnarray}
where ${\textbf{r}_i \equiv (r_i, \theta_i)}$ and ${\boldsymbol \rho_i \equiv ( \rho_i, \theta_i ) = ( \epsilon r_i, \theta_i )}$ are unstretched and stretched radius vectors in the frame of reference co-moving with the $i$-th drop, respectively (Figure~\ref{fig_2drops}). In what follows we will show that interaction of a pair of distant droplets is encapsulated within $N_i^{(2)}$, while terms $N_i^{(0)}$, $N_i^{(1)}$, $F_i^{(1)}$, and $F_i^{(2)}$ may be computed for each droplet individually.

\subsection{Problems at $\epsilon^0$ and $\epsilon^1$: isolated drops}
\label{asymp_o0o1}
To compute the terms $N_i^{(0)}$, $N_i^{(1)}$, and $F_i^{(1)}$ of expansions~\eqref{asymp_n_exp}--\eqref{asymp_f_exp}, we consider each of the droplets separately. That is, we substitute Eqs.~\eqref{asymp_n_exp}--\eqref{asymp_f_exp} into the problem formulated by Eqs.~\eqref{diff-adv}--\eqref{forcefree} in the case of a solitary spherical active drop. We then collect $O(1)$ terms and recover the isotropic solution,
\begin{equation}
  \label{asymp_n0}
  N_i^{(0)}(\textbf{r}_i) = 1 / r_i,
\end{equation}
that corresponds to a motionless droplet and quiescent fluid.

We proceed with the solution of the problem formulated by the $O(\epsilon)$ terms. Since the flow field does not vanish at $\epsilon^1$, we need to write a solution to Stokes equations~\eqref{StokesEq} within and around a spherical drop. In contrast to advection-diffusion equation that requires a composite solution, Eqs.~\eqref{asymp_n_exp}--\eqref{asymp_f_exp}, Stokes equations are linear and admit a solution that is uniformly valid. At each order in $\epsilon$, the axisymmetric solution of the Stokes problem, Eq.~\eqref{StokesEq}, within and outside of the spherical droplet is given by a superposition of orthogonal modes~\citep{Lamb45, Happel83, Leal07},
\begin{align}
  \label{asymp_psi_i}
  & \psi_n^i(\textbf{r}) = r^{n+1} \left( 1 - r^2 \right)
      \left( 1 - \mu^2 \right) L_n'(\mu), \\
  \label{asymp_psi_o}
  & \psi_n^o(\textbf{r}) = \begin{cases}
    \left( 1 - r^3 \right) \left( 1 - \mu^2 \right) / r \quad & n = 1\\
    \left( 1 - r^2 \right) 
      \left( 1 - \mu^2 \right) L_n'(\mu) / r^n \quad & n > 1
  \end{cases},
\end{align}
where $\psi^{i,o}_{n}$ denote the streamfunctions corresponding to the $n$-th mode of the flow decomposition within and outside of the drop, respectively.

Equation~\eqref{asymp_psi_o} implies that in the reference frame co-moving with the drop, the far-field flow at $O(\epsilon)$ is unidirectional and the far-field advection-diffusion equation for each droplet may thus be rewritten as,
\begin{equation}
  \label{asymp_o1_ad}
  -\pe \boldsymbol V_i^{(1)}
      \cdot \boldsymbol \nabla_{\boldsymbol \rho} F_i^{(1)}(\boldsymbol \rho_i)
  = \nabla_{\boldsymbol \rho}^2 F_i^{(1)}(\boldsymbol \rho_i),
\end{equation}
where $\boldsymbol \nabla_{\boldsymbol \rho}$ denotes the gradient in stretched coordinates. Since we disregard the droplet interactions at this order, we may adopt the corresponding solutions for $N_i^{(1)}$, and $F_i^{(1)}$ obtained by~\citet{Morozov19a, Morozov19b},
\begin{align}
  & N_i^{(1)}(\textbf{r}_i) 
    = -2 V_i^{(1)} \left( 1 + \mu_i + \mu_i \frac{2-3r_i}{4 r_i^3} \right), \quad
  & F_i^{(1)}(\boldsymbol \rho_i) 
    = \frac{e^{-2 V_i^{(1)} \rho_i (1+\mu_i)}}{\rho_i},
\end{align}
where ${V_i^{(1)} = \big| \boldsymbol V_i^{(1)} \big|}$ and ${\mu_i \equiv \cos \theta_i}$. Note this effectively implies that $V^{(1)}_i>0$ and that we define $\theta_i$ (and therefore $\mu_i$) from an axis of reference that is oriented along the direction of propulsion of each droplet, i.e. toward (resp. away from) the second droplet for the approaching (resp. departing) case (Figure~\ref{fig_2drops}).

Also note that Eqs.~\eqref{asymp_n_exp}, \eqref{asymp_psi_o} and~\eqref{asymp_n0} indicate that in the lab frame the leading order flow at a distance $1/\epsilon$ from an active droplet is $O(\epsilon^4)$ (recall that at $\epsilon^0$ the fluid is motionless), while the droplet's chemical footprint is $O(\epsilon)$. That is, at a distance $1/\epsilon$ from  both  droplets, the advection diffusion equation, Eq.~\eqref{diff-adv}, reduces to an unsteady diffusion equation in the lab frame,
\begin{equation}
  \label{asymp_ad_lab}
  \pe \frac{\partial F}{\partial t} = \nabla^2 F + O(\epsilon^4),
\end{equation}
where $F$ denotes the combined far-field concentration footprint of a pair of active drops, time derivative accounts for the displacement of the droplets in the lab frame, and $O(\epsilon^4)$ corresponds to the contribution of the flow field. Equation~\eqref{asymp_ad_lab} is linear, therefore $F$ can be found as a superposition of the contributions from the individual droplets,
\begin{equation}
  \label{asymp_super}
  F = F_1 + F_2
    = \epsilon \left( F_1^{(1)} + F_2^{(1)} \right)
    + \epsilon^2 \left( F_1^{(2)} + F_2^{(2)} \right)
    + \ldots.
\end{equation}

Finally, we demonstrate that interaction between the droplets appears only in the problem at $O(\epsilon^2)$. To this end, we write the concentration field of droplet 2 in the coordinate system of droplet 1 and expand the result in powers of $\epsilon$ in the cases of approaching and departing drops, respectively,
\begin{align}
  \label{asymp_inter}
  & F_{2,\text{approach}} = \epsilon \frac{e^{-8 D V^{(1)}}}{2D}
    - \epsilon^2 \frac{e^{-8 D V^{(1)}} \left( 1 + 8 D V^{(1)} \right)}{4D^2} r_1 \mu_1
    + O( \epsilon^3 ), \\
  \label{asymp_dep}
  & F_{2,\text{departure}} = \epsilon \frac{1}{2D}
    + \epsilon^2 \frac{r_1 \mu_1}{4D^2}
    + O( \epsilon^3 ),
\end{align}
where $V^{(1)} \equiv V_1^{(1)} = V_2^{(1)}$. Note that the first term in the right-hand side of Eqs.~\eqref{asymp_inter}--\eqref{asymp_dep} is constant and, thus, can not produce any Marangoni stresses. In turn, the second term in the right-hand side of Eq.~\eqref{asymp_inter} constitutes a unidirectional concentration gradient that implements the effect of droplet 2 onto droplet 1. This term is quadratic in $\epsilon$, which justifies that the droplets may be considered separately at $O(\epsilon^0)$ and $O(\epsilon^1)$.

\subsection{Problem at $\epsilon^2$: droplet interaction}
Section~\ref{asymp_o0o1} established that the interaction of a pair of distant droplets is implemented by a linear concentration gradient in~\eqref{asymp_inter}. On the other hand, in the reference frame co-moving with the drop, the near-field advection-diffusion, Eq.~\eqref{diff-adv}, reduces to an inhomogeneous steady diffusion equation~\citep{Morozov19a, Morozov19b} that admits Eq.~\eqref{asymp_inter} as a solution. Therefore, the influence of the concentration gradient imposed by droplet 2 in the vicinity of droplet 1 only appears in the boundary conditions for droplet 1 and vice versa. As a result, general solutions for $N_i^{(2)}$, and $F_i^{(2)}$ may be adapted directly from Refs.~\citep{Morozov19a, Morozov19b},
\begin{multline}
  N_i^{(2)} (\textbf{r}_i)
  = \frac{C_{0,i}}{r_i}
  + \frac{(V^{(1)})^2}{30 r_i^5} \left( 8 - 15 r_i + 20 r_i^3 + 80 r_i^6 \right)
  - \frac{1}{2} \left( \delta V^{(1)} + 8 A_{1,i} \right)\\
  + L_1(\mu_i) \left( 
      \frac{C_{1,i}}{r_i^2} + 4 r_i (V^{(1)})^2
      - \frac{1 + 2 r_i^3}{4 r_i^3} \left( \delta V^{(1)} + 8 A_{1,i} \right)
    \right) \\
  + L_2(\mu_i) \left(
      \frac{C_{2,i}}{r_i^3}
      + \frac{(V^{(1)})^2}{21 r_i^5} 
          \left( 10 - 21 r_i + 70 r_i^3 - 42 r_i^4 + 28 r_i^6 \right)
      - \frac{4 + 6 r_i^2}{r_i^4} A_{2,i}        
    \right),
\end{multline}
\begin{equation}
  F_i^{(2)}(\boldsymbol \rho_i) 
    = -\frac{1+\mu_i}{2} e^{-2 V^{(1)} \rho_i (1+\mu_i)} \left( \delta V^{(1)} + 8 A_{1,i} \right),
\end{equation}
where $A_{1,i}$, $A_{2,i}$, $C_{0,i}$, $C_{1,i}$, and $C_{2,i}$ are unknown coefficients to be determined from the boundary conditions. Following the procedure described in Ref.~\citep{Morozov19a}, we substitute $N_i^{(2)}(\textbf{r}_i)$ and $F_i^{(2)}(\boldsymbol \rho_i)$ into the boundary conditions at $\epsilon^2$ and obtain a set of algebraic equations in terms of the coefficients $A_{1,i}$, $A_{2,i}$, $C_{0,i}$, $C_{1,i}$, and $C_{2,i}$. Solvability condition of this set of equations yields the leading-order self-propulsion velocity of the droplet exposed to a weak concentration field,
\begin{equation}
  \label{asymp_sol_uinf}
  V^{(1)} = \frac{\delta \pm \sqrt{\delta^2 + 256 G}}{32},
\end{equation}
where 
\begin{equation}
  G = G_\text{approach} \equiv -\frac{e^{-8 D V^{(1)}} \left( 1 + 8 D V^{(1)} \right)}{4 D^2}
  \quad \text{or} \quad
  G = G_\text{departure} \equiv \frac{1}{4 D^2}
\end{equation}
for approaching and departing drops, respectively. We emphasise that ${V^{(1)} \geq 0}$ by definition. As a result, the departing configuration for which $G>0$ is used in Eq.~\eqref{asymp_sol_uinf} always admits a single solution, while the approaching configuration for which $G<0$ may have zero or two solutions depending on the magnitude of the concentration gradient. 

\begin{figure}
  \centering
  \includegraphics[width=.6\textwidth]{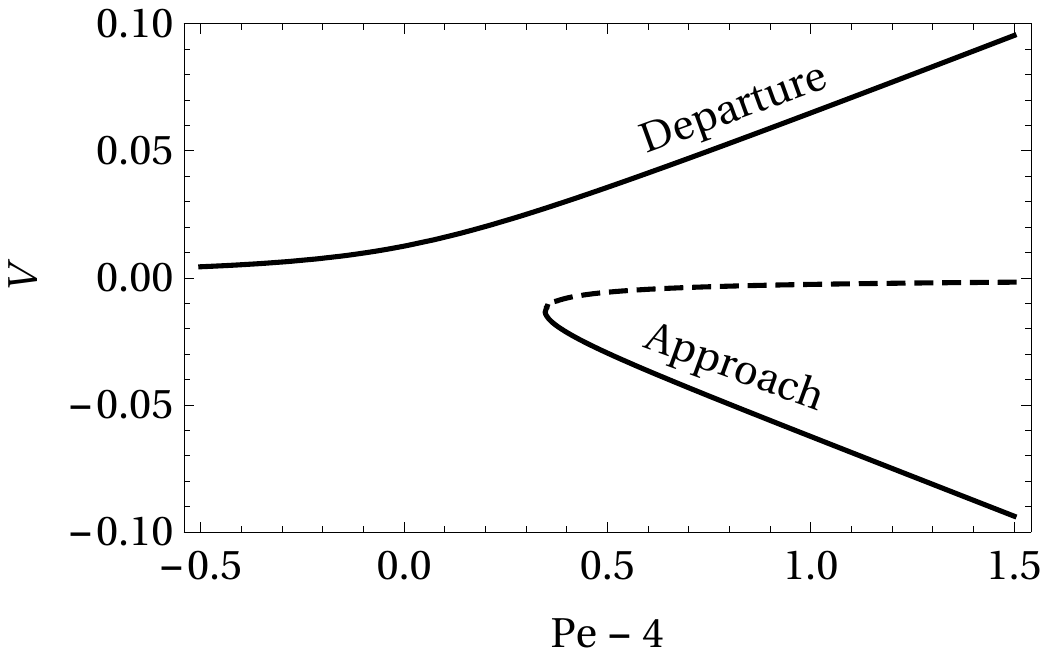}
  \caption{
    Bifurcation diagram~\eqref{asymp_sol_uinf}: 
    leading-order self-propulsion velocities 
    of a pair of identical active droplets located at a distance $2d=2D/\epsilon$ with $D = 10$.
    Top and bottom branches correspond 
    to departing and approaching drops, respectively.
    A solid (resp. dashed) line denotes a stable (resp. unstable) propulsion regime.
  }
  \label{asymp_bif_diag}
\end{figure}

Reconstructing the absolute relative velocity of the droplets is achieved as ${V=-\epsilon V^{(1)}_\textrm{approach}}$ (resp. ${V=\epsilon V^{(1)}_\textrm{departure}}$) for the approaching (resp. departing) case. In our numerical analysis, droplet approach corresponds to negative velocity (as shown in \change{Figures~\ref{collision_overviewPe6} and \ref{collision_overviewPe20}, \ref{early_chemicalPe20} and \ref{collision_overview_drop}}), and the bifurcation diagram~\eqref{asymp_sol_uinf} is plotted in Figure~\ref{asymp_bif_diag} to match this convention. Equation~\eqref{asymp_sol_uinf} corresponds to an imperfect transcritical bifurcation implying that not all of the branches of the bifurcation diagram~\eqref{asymp_sol_uinf} are stable, as shown in Figure~\ref{asymp_bif_diag}. Also note that the bifurcation diagram corresponding to Eq.~\eqref{asymp_sol_uinf} is quasistatic, i.e., it depends on time through the separation distance between the droplets, $D$, only.

In essence, Eq.~\eqref{asymp_sol_uinf} establishes that the regime of steadily approaching droplets does not exist when ${G_\text{approach} < G_c \equiv - \delta^2 / 256<0}$. As the droplets approach each other, $G$ is negative and increases in magnitude as $D$ decreases, up to a point where the quasi-steady approach branch ceases to exist, which is identified to the droplet's velocity reversal and rebound dynamics. This event is associated with a critical center-to-center distance $\tilde D/\epsilon=\tilde d_c$ which provides an estimate for the rebound distance $d_\textrm{rb}$ and satisfies
\begin{equation}
  \label{asymp_sol_uinf_drop}
  \left( \tilde D \delta / 4 \right)^2 
    = 4 \left( 1 + \tilde D \delta / 4 \right) e^{-\tilde D \delta / 4}
\end{equation}
and
\begin{equation}
  \label{asymp_sol_uinf_drop_final}
  \tilde d_\text{c} \approx \frac{5.98}{\pe - 4}.
\end{equation}
Estimation of the droplet rebound distance~\eqref{asymp_sol_uinf_drop_final} reproduces the scaling obtained from our numerical simulations (Section~\ref{SeveralPe}). The prefactor of this asymptotic estimate $\tilde d_c$ of the rebound distance differs however slightly from the value estimated form our numerical results (\change{Figure~\ref{rebound_distance}}). Such discrepancy is to be expected, since in numerical simulations rearrangement of the concentration field in the course of the rebound is not immediate, and in fact even in the quasi-steady framework presented here, $|V|=V_0/2\neq 0$ when the approaching branch ceases to exist (Figure~\ref{asymp_bif_diag}). Instead, droplets take some time to slow down. This regime of transitory approach is beyond reach of our asymptotic analysis and, thus, the real rebound distance is shorter, compared to the theoretical prediction, Eq.~\eqref{asymp_sol_uinf_drop_final}.

The analysis presented here is focused on the case of two symmetric droplets. Yet, as emphasised throughout the analysis the coupling between the droplets is purely chemical. Indeed, the flow field contribution is limited to the near-field dynamics of the flow field and does not influence the far-field signature (only the displacement of the droplet does). This further validates that, near the self-propulsion threshold, hydrodynamic interactions play a subdominant role. While chemical interactions are mediated through chemical gradients which decay as $1/d^2$, direct hydrodynamic interactions (i.e. the drift of a droplet in the flow field of the second one) would be dominated by the stresslet flow created by each droplet, which also decays as $1/d^2$. However, the intensity of the stresslet itself is weak for $\pe\approx\pe_c$, namely scaling as $\epsilon^2$~\citep{Morozov19a}, so that hydrodynamic interactions are $O(\epsilon^4)$ in contrast with $O(\epsilon^2)$ chemical interactions. As a consequence, the present asymptotic analysis applies exactly to the case of a droplet collision with a no-slip wall.\\

Finally, we note that the present approach and bifurcation diagram in Eq.~\eqref{asymp_sol_uinf} applies to any active droplet exposed to a concentration field, $C_e$, that allows for an expansion in powers of $\epsilon$ as shown in Eqs.~\eqref{asymp_inter}--\eqref{asymp_dep}. The physical meaning of this mathematical requirement is twofold: (i) the evolution of $C_e$ must be slow, that is, in the lab frame $C_e$ should satisfy the steady diffusion equation up to $O( \epsilon^3 )$,
\begin{equation}
  \label{asymp_ce_req1}
  \nabla^2 C_e = O( \epsilon^3 ),
\end{equation}
and (ii) the gradient of $C_e$ must be weak, namely, $C_e$ may only contribute to $N_i^{(2)}$, that is,
\begin{equation}
  \label{asymp_ce_req2}
  \boldsymbol \nabla C_e = O( \epsilon^2 ).
\end{equation}
Any $C_e$ that satisfies these requirements, Eqs.~\eqref{asymp_ce_req1}-\eqref{asymp_ce_req2}, may be seamlessly included in the superposition in Eq.~\eqref{asymp_super} and subsequently expanded in powers of $\epsilon$ to obtain the corresponding value of $G$ for the bifurcation diagram~\eqref{asymp_sol_uinf}.

\section{Effective interactions}
\label{Effective-interaction}

The results of \S~\ref{DropletInteraction} emphasised the complexity of the interaction of the droplet with the confining wall (or with a second droplet) and the diversity of detailed behaviour when varying $\pe$. These results provide significant insight into such interactions and collisions that we may wish to implement in the modelling of more complex systems where full treatment of the coupled chemical and hydrodynamic problems is not achievable anymore. This includes, for example, the dynamics of a large number of droplets as observed experimentally, where the dynamics of each droplet can be seen as the succession of self-propelling stages (i.e. isolated dynamics) and collisions with neighbours and/or boundaries. The purpose of the present section is therefore to provide a global effective characterisation of the rebound. 

The results of \S~\ref{DropletInteraction} show that each collision is not simply the sequence of a self-propulsion with $\bs V=-V_0\bs e_z$ toward the wall followed by a propulsion one at $\bs V=V_0\bs e_z$ away from it. Indeed, the droplet may experience gradual slowdowns or a velocity plateau and may rebound at a different distance $d_\textrm{rb}$ depending on the exact ratio of diffusion and advection of the solute as quantified by $\pe$. However, the initial and final stages of the sequence are always the same, namely propulsion with velocity $\pm V_0\eb_z$, so that the main quantity of interest when looking at the long-term dynamics is the total duration of the collision, or equivalently the excess time taken in comparison with an elastic shock (i.e. where a droplet would self-propel constantly at $\pm V_0\eb_z$ and rebound on the wall). 

\begin{figure}
\centering
\includegraphics[width=.7\textwidth]{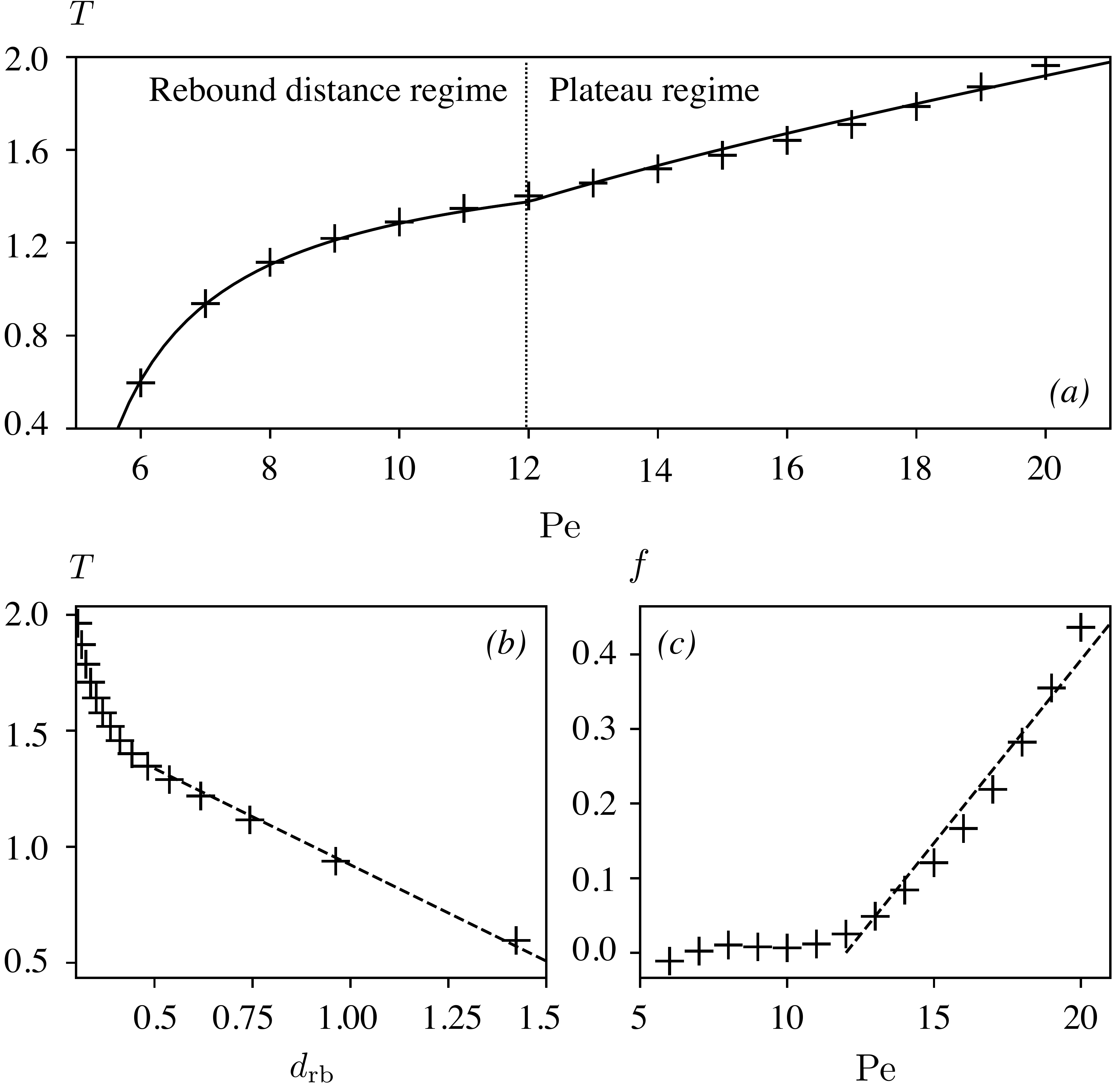}
\caption{Time needed for the droplet to travel from a distance $d=d_m$ from the wall and coming back after collision at the same distance. $(a)$:  Ratio of the time travel $\Delta t$ to the time $2d_m/V_0$ corresponding to the time taken by a particle travelling at velocity $V_0$ from $d=d_m$ to the same point after a rigid collision with the wall. $(b)$: Extra time taken by the droplet compared to the rigid collision case as function of the rebound distance $d_{\text{rb}}$. $(c)$: $f$ evolution as function of Pe. }
\label{effective_model}
\end{figure}

We attempt to characterise here this quantity, and therefore the collision, using the following protocol:  considering a (large) reference distance $d_m$ away from the wall, we measure the corresponding time $\Delta t$ needed for an active droplet to travel from $d=d_m$ towards the wall, rebound and come back at the same location. This lapse of time is then compared to $2d_m/V_0$, which is the time taken by a particle moving at the constant velocity $-V_0\eb_z$ and experiencing a rigid collision \emph{on the wall} before returning to its original position with constant velocity $V_0\eb_z$. Their difference is the delay introduced by the full hydro-chemical dynamics with respect to a simple elastic shock, and we thus define the \emph{relative excess collision time} as 
\begin{equation}
T=\frac{V_0\Delta t-2d_m}{R}.
\end{equation}
The variations of $T(\pe)$ are shown on Figure~\ref{effective_model}(a). First, one should observe that, within the range of $\pe$ explored here, the relative collision time, $T$, is positive, meaning that the collision of a self-propelled droplet takes always more time than the rigid particle collision. This is a result of two competing effects, the rebound of the droplet at a finite distance away from the wall (i.e. it actually travels a distance shorter than $2d_m$) and its slowed-down dynamics in the vicinity of the wall, and $T>0$ suggests that the latter is dominant. Secondly, two different regimes can be identified: for moderate $\pe$ and up to $\pe\approx 12$ the relative collision time $T$ evolves concavely whereas it is mostly linear for higher $\pe$.

A better understanding of the origin of these two regimes stems from  two main phenomena that cause an increase of $\Delta t$ (and $T$) with $\pe$. As $\pe$ increases away from $\pe_c$, the rebound distance $d_{\text{rb}}$ decreases and the droplet travels a longer distance before coming back.  A particle propelling at velocity $\pm V_0$ and bouncing back at a distance $d_{\text{rb}}$ from the wall would take a time $2(d_m-d_{\text{rb}})/V_0$ before returning to its initial position, and the corresponding relative excess collision time would be $-2d_{\text{rb}}/R$. The asymptotic linear dependence between $T$ and $d_\textrm{rb}$, observed in Figure~\ref{effective_model}(b), therefore suggests that the increase in relative collision time $T$ for the active droplet is caused by the change of $d_{\text{rb}}$ with $\pe$. This argument only explains the increase with $\pe$ rather than the absolute variations: indeed, the droplet's velocity magnitude is smaller than $V_0$ for a significant part of the sequence so that $T>0$ while $T<0$ for the particle rigid collision at $d_\textrm{rb}$. Nevertheless, these observations suggest variations of $T$ for moderate $\pe$ of the form:
\begin{eqnarray}
T=\frac{\Delta t V_0-2d_m}{R}= K_1+K_2d_\textrm{rb},\label{time_pred}
\end{eqnarray}
Using a Gauss-Newton method, the best least-square fit for $K_1$ and $K_2$ is obtained as $K_1=1.75$ and $K_2=-0.83$. 

Figure~\ref{rebound_distance} shows that the decrease of $d_{\text{rb}}$ is less pronounced for higher $\pe$, which suggests that a second phenomenon is responsible for the increase in relative collision time $T$ at higher $\pe$. As emphasised in Section~\ref{DropletInteraction}, a distinctive feature of the larger $\pe$ collisions is the development of a velocity plateau during which the droplet maintains a rather constant velocity smaller than $V_0$ after rebounding on the wall. This plateau is  $\pe$-dependent and Figure~\ref{effective_model}(c) shows the evolution of the function $f$ defined by the difference at larger $\pe$ between the actual value of $T$ and its prediction of Eq.~\eqref{time_pred}:
\begin{eqnarray}
f=T-K_1-K_2\frac{d_{\text{rb}}}{R}.
\end{eqnarray}
Figure~\ref{effective_model}(c) shows that $f$ is reasonably well approximated by a linear function of $\pe$, so that a global effective model for the relative collision time is obtained as:
\begin{eqnarray}
\label{effectiveformula}
T=\frac{\Delta t V_0-2d_m}{R}=&=&K_1+K_2\frac{d_{\text{rb}}(\mbox{Pe})}{R}+K_3\,\text{max}(0,\mbox{Pe}-12),
\end{eqnarray}
where the coefficient $K_3=0.05$ is fitted through Gauss-Newton non-linear regression method. Obtaining an effective model finally requires an expression of $d_\textrm{rb}$ as a function of $\pe$.  Inspired by the asymptotic approach of section \ref{asymp}, a simple model is chosen of the form:
\begin{eqnarray}
d_{\text{rb}}(\mbox{Pe})&=&\frac{K_4}{\mbox{Pe}-4}+\frac{K_5}{\sqrt{\pe-4}},
\end{eqnarray}
with fitted constants $K_4=1.89$ and $K_5=0.61$ determined from the data of Figure~\ref{rebound_distance}. 

The resulting effective model for the excess relative collision time $T$ is shown on Figure~\ref{effective_model}(a) as a solid black line, and appears to provide a reasonable estimate of the collision time $T$ for the range of $\pe$ investigated here (i.e. $\pe\in[6\,,\,20]$). It includes the two main physical features of the collision dynamics for varying $\pe$, namely the change in rebound distance and the existence of a velocity plateau.

\section{Conclusions}\label{conclusion}
The present work provides a unique insight into the interaction and rebound dynamics of a chemically-active droplet with a rigid confining wall (as well as the related problem of the symmetric collision of two such droplets). In contrast with most existing studies that rely on some assumptions regarding either the simplified solute transport or the relative distance to the wall, the unsteady dynamics of the solute concentration and its coupling to the hydrodynamic here are fully resolved here for any relative distance. This provides a quantitative analysis of the detailed solute transport around the droplet during its rebound. 

In particular, we show that for moderate $\pe$, namely the ratio of convective and diffusive solute transport, the rebound dynamics is well-captured by neglecting the hydrodynamic effect of the wall and can be understood as the slowdown, reversal and re-acceleration of the droplet in an adverse chemical gradient whose magnitude increases as the relative distance is decreased. In contrast, when advection becomes more dominant, the complex hydrodynamic flow around the confined droplet imposes a reorganisation of the chemical field that profoundly alters its swimming and rebound dynamics, with a significant reduction in the minimum distance to the wall  and the emergence of a velocity plateau after the rebound, during which the droplet maintains a reduced and somewhat constant velocity before accelerating again to its nominal value as it escapes the region of influence of the wall. This phenomenon can be related to the self-sustained gradients in surface solute concentration by the Marangoni flows they generate, even when the droplet is forced to slow down and stop by the hydrodynamic effect of the wall.

To retain the relative simplicity of an axisymmetric problem, the configuration considered here is highly symmetric as only the normal approach of a single droplet to a flat wall is considered. Yet, this provides an important physical insight into the interaction and rebound dynamics, which could contribute significantly to a better understanding of experimental studies involving confined active droplets: several recent contributions have indeed suggested that the collective behaviour of many self-propelled droplets is greatly influenced by the role of confinement on their interactions~\citep{kruger2016dimensionality,thutupalli2018flow}. The present analysis also provides a critically-valuable benchmark analysis for the validation of simpler models (e.g. relying on far-field approximations or simplified interactions) that could be used for more complex problems (non-normal rebound or interactions of many droplets).

In the vicinity of the self-propulsion threshold, $\pe\sim \pe_c$, weakly-nonlinear theory of the droplet-droplet interaction confirms that the flow field created around a given droplet by the presence of the wall or another droplet is negligible and that the coupling is purely chemical (\S~\ref{asymp}). Thereby, it rigorously establishes that the symmetric collision of two droplets and the rebound on a rigid wall or free-surface are equivalent at leading order in that limit.  The slowdown, and eventual rebound dynamics, are then interpreted in the framework of imperfect transcritical bifurcations as the disappearing of one of the stable solution branches, corresponding to the propulsion of the droplet against a steepening chemical gradient (Figure~\ref{asymp_bif_diag}). Such an event occurs for distances that scale as $d\sim 1/(\pe-\pe_c)$, which is validated against our numerical solution of the full problem (\change{Figure~\ref{rebound_distance}}). We further demonstrate that due to the purely chemical nature of weak droplet-droplet interactions, this framework and the resulting bifurcation diagram, Fig.~\ref{asymp_bif_diag}, applies to any active droplet exposed to an externally-imposed spatially-evolving solute concentration $C_e(\mathrm{r})$, provided the variations of $C_e$ are slow enough on the scale of the droplet, Eqs.~\eqref{asymp_ce_req1}--\eqref{asymp_ce_req2}. As such, the conclusions of our weakly-nonlinear analysis capture a universal feature of active droplet dynamics.

From a more technical point of view, the numerical approach followed here provides a novel framework for the simulation and spectral analysis of time-dependent problems in a bi-spherical geometry. At each instant, a coordinate system is used that fits the natural boundaries of the problem which is particularly well-suited for time-dependent multi-physics problems where two different dynamics are coupled on the moving boundary (here the hydrodynamic flow field and the solute concentration).  In this work, we use this framework to analyse two geometrically-simple problems (i.e. a single droplet and a flat wall or two identical droplets), yet, it can be straightforwardly used to treat more complex situations such as the rebound on a curved wall or droplets of different sizes. Such generalisations are beyond the scope of the present paper, but deserve a more complete treatment. In particular, phoretic particles are known to exhibit non-reciprocal interactions~\citep{michelin2015autophoretic,soto2014,soto2015self} which stem from the coupling of two distinct \change{physico-chemical} properties (activity and mobility) to generate self-propulsion and that can result in complex dynamics when coupling particles of different nature or sizes. A similar property can thus be expected for active droplets since their self-propulsion also rely on this activity-mobility combination.

 Using bi-spherical coordinates to solve diffusion or viscous flow problems is obviously not new~\citep{stimson1926motion,popescu2011,michelin2015autophoretic,michelin2019bubble} but has so far been limited to quasi-steady problems where Laplace or Stokes equations are solved independently at each instant. In contrast here, the advection-diffusion dynamics requires accounting for the non-trivial evolution of the grid. This is particularly useful for active droplets whose underlying physics critically relies on the non-linearity introduced by the advection-diffusion of the solute. Yet, it may also prove particularly useful to analyse a variety of other time-dependent problems such as the unsteady mass transfer and viscous growth/dissolution of gas bubbles (e.g. near catalytic surfaces or during boiling), or the collective dynamics of such bubbles or droplets~\citep{michelin2018bubble}.

\section*{Acknowledgments}
This project has received funding from the
European Research Council (ERC) under the European Union's Horizon
2020 research and innovation programme under Grant Agreement 714027 (SM).

\section*{Declaration of Interests}
The authors report no conflict of interest.

\appendix

\section{Modal decomposition and projection in the Legendre basis}
\label{OperatorProjection}
\subsection{Projection of the advection-diffusion equation, Eq.~\eqref{modal_adv_diff}}
The advection-diffusion equation, Eq.~\eqref{modal_adv_diff}, is projected along the $p$-th Legendre polynomials by integrating over $\mu$ its product with $L_p(\mu)$. Using the generating function of the Legendre polynomials, one can express
\begin{equation}\label{genfun}
\frac{1}{\Gamma^{1/2}}=\frac{1}{\sqrt{\cosh(\lambda\xi)-\mu}}=\sqrt{2}\sum_{k=0}^\infty L_k(\mu)\mathrm{e}^{-(p+1/2)|\lambda\xi|},
\end{equation}
and each term of the projected version of Eq.~\eqref{mainPDE} can be obtained as:
\begingroup
\allowdisplaybreaks
 \begin{align}
H_{pn}&=\sqrt{2}\sum_{k=0}^{\infty}Q^0_{knp}e^{-(k+1/2)|\lambda\xi|},\\
G^1_{pn}&=\frac{\dot{a}}{a\sqrt{2}}\sum_{k=0}^{\infty}\left(Q^0_{knp}+\cosh(\lambda\xi) Q^1_{knp}-2\cosh(\lambda\xi)R^0_{knp}   \right)e^{-(k+1/2)|\lambda\xi|},\\
G^2_{pn}&=\sqrt{2}\sum_{k=0}^{\infty}\left(\frac{\dot{a}\sinh(\lambda\xi)}{a\lambda}Q^1_{knp}-\frac{\dot{\lambda}\xi}{\lambda}Q^0_{knp}\right)e^{-(k+1/2)|\lambda\xi|},\\
B^1_{pnk}&=\frac{1}{a^3}\left(\frac{3\sinh(\lambda\xi)}{2}S^0_{knp}-\frac{k(k+1)}{2}\sinh(\lambda\xi)Q^0_{knp}\right),\\
B^2_{pnk}&=\frac{1}{\lambda a^3}\left(-\cosh(\lambda\xi)S^0_{knp}+S^1_{knp}+\frac{1}{2}R^0_{nkp}\right),\\
B^3_{pnk}&=\frac{1}{\lambda a^3}\left(\frac{3}{2}R^0_{nkp}-k(k+1)\left(\cosh(\lambda\xi)Q^0_{knp}-Q^1_{knp}\right) \right),\\
A^2_{pn}&=\frac{\sqrt{2}}{\lambda^2 a^2}\sum_{k=0}^\infty\left(\cosh^2(\lambda\xi)Q^0_{knp}-2\cosh(\lambda\xi)Q^1_{knp}+Q^2_{knp}\right)e^{-(k+1/2)|\lambda\xi|},\\
A^1_{pn}&=-\lambda^2\left(n+\frac{1}{2}\right)^2A_{pn}^2.
 \end{align}
 \endgroup
where $Q_{knp}^i$, $R^i_{knp}$ and $S^i_{knp}$ are the following integrals of Legendre polynomials 
\begin{align}
\label{Qdef}
Q_{knp}^i&=\int_{-1}^1\mu^i L_n L_k L_p  \dd \mu,\\
\label{Rdef}
R_{knp}^i&=\int_{-1}^1\mu^i(1-\mu^2) L_n' L_k L_p  \dd\mu=\frac{n(n+1)}{2n+1}\left(Q^i_{k,n-1,p}-Q^i_{k,n+1,p}\right),\\
\label{Sdef}
S_{knp}^i&=\int_{-1}^1\mu^i(1-\mu^2) L_n' L_k' L_p  d\mu,
\end{align}
and can be obtained recursively using classical relations between Legendre polynomials~\citep{abramowitz1964}:
 \begin{align}
 Q_{0,n,p}^0&=\frac{2}{2p+1}\delta_{n,p},\qquad\qquad\qquad Q_{1,n,p}^0=\frac{n+1}{2n+1}\delta_{p,n+1}+\frac{n}{2n+1}\delta_{p,n-1},\\
  Q_{k,n,p}^0&=\frac{2k-1}{k}\left( \frac{n+1}{2n+1}Q^0_{k-1,n+1,p}+\frac{n}{2n+1}Q^0_{k-1,n-1,p}   \right) -\frac{k-1}{k}Q^0_{k-2,n,p},\\
 Q_{0,n,p}^i&=Q^{i-1}_{1np},\qquad\qquad\qquad\qquad\,\, Q_{k,n,p}^i=\frac{k+1}{2k+1} Q_{k+1,n,p}^{i-1}+\frac{k}{2k+1} Q_{k-1,n,p}^{i-1},\\
 S_{0,n,p}^i&=S_{n,0,p}^i=0,\,\,\qquad\qquad\qquad S_{k,n,p}^0=S_{k-1,n,p}^0+(2k-1)R_{k-1,n,p}^{0},\\
  S_{k,n,p}^i&= S_{k+1,n,p}^{i-1}-(k+1) R_{k,n,p}^{i-1}.\label{eq:Srec}
 \end{align}
 
 \subsection{Projection of the hydrodynamic boundary conditions}
\label{ProjHyd}
Substituting the definition of $\psi^{i,o}$, Eq.~\eqref{psiLegendre}, as well as Eq.~\eqref{genfun} into Eqs.~\eqref{impermeabilitybisph}  and \eqref{continuity2} leads after projection onto $(1-\mu^2)L_n'$:
\begin{align}
\label{BCH1}
\left.U^i_{n}\right|_{\xi=1}=\left.U^o_{n}\right|_{\xi=1}&=\frac{\sqrt{2}Va^2}{2}\left( \frac{e^{-(n-1/2)|\lambda|}}{2n-1}-\frac{e^{-(n+3/2)|\lambda|}}{2n+3}\right),\\
\label{BCH2}
\left.\frac{\partial U^o_{n}}{\partial \xi}\right|_{\xi=1}&=\left.\frac{\partial U^i_{n}}{\partial \xi}\right|_{\xi=1}.
\end{align}
Similarly, at the wall surface, the no-slip conditions simply write
\begin{align}
\label{BCH5}
\left.U_{n,o}\right|_{\xi=0}&=0,\\
\label{BCH6}
\left.\frac{\partial U_{n,o}}{\partial \xi}\right|_{\xi=0}&=0.
\end{align}

The tangential shear stress at the surface of the droplet is obtained as:
\begin{align}
\left.\sigma^{i,o}_{\xi\mu}\right|_{S}&=-\frac{\Gamma^{3/2}}{a^3}\sum_{n=1}^{\infty}\sqrt{1-\mu^2}L_n'\left[\frac{1}{\lambda^2}\frac{\partial^2 U_n^{i,o}}{\partial \xi^2}+\left(n(n+1)-\frac{3}{4}\left(1+\frac{2\sinh(\lambda\xi)^2}{\Gamma^2}  \right)\right)U_n^{i,o}    \right]
\end{align}
and substitution into the Marangoni condition, Eq.~\eqref{marangonibisph}, provides the following condition at the droplet surface ($\xi=1$):
\begin{align}
\sum_{n=1}^{\infty}L_n' \bigg\{\frac{\Gamma^2}{\lambda^2}\left(\frac{\partial^2 U_n^o}{\partial \xi^2}-\tilde{\eta}\frac{\partial^2 U_n^i}{\partial \xi^2}\right)&+\left[\left(n^2+n-\frac{3}{4}\right)\Gamma^2-\frac{3}{2}\sinh^2(\lambda\xi)\right]\left(U_n^o-\tilde{\eta}U_n^i\right)\left.\bigg\}\right|_{\xi=1}\nonumber\\
&=-(2+3\tilde{\eta})a^2\sum_{n=0}^{\infty}\left.\left[c_n\left( -\frac{\Gamma L_n}{2}+\Gamma^2 L_n'\right)\right]\right|_{\xi=1}\label{marangonimodes}
\end{align}
Projecting the previous equation onto $(1-\mu^2)L'_p(\mu)$  finally leads to:
\begin{align}
\sum_{n=1}^{\infty}&\frac{\bar{S}_{np}(\lambda)}{\lambda^2}\left.\left[\frac{\partial^2 U_n^o}{\partial \xi^2}-\tilde{\eta}\frac{\partial^2 U_n^i}{\partial \xi^2}+\lambda^2\left(n^2+n-\frac{3}{4}\right)\left(U_n^o-\tilde{\eta}U_n^{i}\right)\right]\right|_{\xi=1}\nonumber\\
-&\frac{3p(p+1)\sinh^2\lambda}{2p+1}\left.\left(U_p^o-\tilde{\eta}U_p^{i}\right)\right|_{\xi=1}=(2+3\tilde{\eta})a^2\sum_{n=0}^{\infty}\left[\frac{\bar{R}_{np}(\lambda)}{2}-\bar{S}_{np}(\lambda)\right]c_n|_{\xi=1},
\end{align}
where the functions $\bar{S}_{np}(\lambda)$ and $\bar{R}_{np}(\lambda)$ are computed from the different integrals in Eqs.~\eqref{Qdef}--\eqref{eq:Srec} as
\begin{equation}
\bar{S}_{np}(\lambda)=S^0_{np0}\cosh^2\lambda-2S^1_{np0}\cosh\lambda +S^2_{np0},\qquad \bar{R}_{np}(\lambda)=R^0_{np0}\cosh\lambda-R^1_{np0}.
\end{equation}

\end{document}